\documentclass[12pt,letterpaper]{article}

\usepackage{amsfonts,amssymb,amsmath,amscd}
\usepackage[ps,dvips,matrix,arrow,frame,import,curve,color]{xy}
\newcommand{\ra}{\rightarrow}

\newcommand{\Tr}{{\rm Tr}}

\newcommand{\ZZ}{{\mathbb Z}}
\newcommand{\RR}{{\mathbb R}}
\newcommand{\CC}{{\mathbb C}}

\newcommand{\cL}{{\mathcal L}}
\newcommand{\cO}{{\mathcal O}}

\newcommand{\be}{{\beta}}

\renewcommand{\part}{\partial}

\newcommand{\bpartial}{{\bar\partial}}

\newcommand{\bpsi}{{\bar\psi}}
\newcommand{\bchi}{{\bar\chi}}

\newcommand{\blambda}{{\bar\lambda}}

\newcommand{\bz}{{\bar z}}
\newcommand{\bw}{{\bar w}}

\newcommand{\ot}{\otimes}
\newcommand{\eps}{\epsilon}

\newcommand{\tq}{{\tilde q}}

\newcommand{\cA}{{\mathcal A}}
\newcommand{\cF}{{\mathcal F}}
\newcommand{\cN}{{\mathcal N}}
\newcommand{\MH}{{{\mathcal M}_H}}
\newcommand{\cE}{{\mathcal E}}

\renewcommand{\Re}{{\rm Re}}

\newcommand{\cD}{{\mathcal D}}
\newcommand{\cR}{{\mathcal R}}

\newcommand{\ad}{{\rm ad}}

\newcommand{\cW}{{\mathcal W}}

\newcommand{\LG}{{{}^LG}}
\newcommand{\PP}{{\mathbb P}}

\newcommand{\cM}{{\mathcal M}}

\newcommand{\cC}{{\mathcal C}}

\newcommand{\vphi}{\varphi}
\newcommand{\bK}{{\bar K}}
\newcommand{\Lt}{{{}^Lt}}
\newcommand{\frg}{{\mathfrak g}}
\newcommand{\dL}{{\widehat\Lambda}}
\newcommand{\sA}{{\mathsf A}}
\newcommand{\sB}{{\mathsf B}}

\newcommand{\vol}{{\rm vol}}
\newcommand{\sV}{{\mathsf V}}

\newcommand{\sU}{{\mathsf W}}
\newcommand{\Hom}{{\rm Hom}}

\newcommand{\frA}{{\mathfrak A}}
\newcommand{\frM}{{\mathfrak M}}
\newcommand{\frV}{{\mathfrak H}}
\newcommand{\frR}{{\mathfrak R}}
\newcommand{\frt}{{\mathfrak t}}
\newcommand{\LR}{{{}^LR}}
\newcommand{\cP}{{\mathcal P}}

\newcommand{\diag}{{\rm diag}}
\newcommand{\cQ}{{\mathcal Q}}
\newcommand{\NN}{{\mathbb N}}
\newcommand{\WW}{{\mathbb W}}
\newcommand{\WP}{{\WW\PP}}

\def\be{\begin{equation}}
\def\ee{\end{equation}}
\def\bear{\begin{eqnarray}}
\def\eear{\end{eqnarray}}

\def\half{{ \frac{1}{2} }}

\def\dg{{\dagger}}
\def\wdg{{\wedge}}
\def\Re{{\rm Re\hskip0.1em}}

\def\dg{{\dagger}}

\def\tC{{{\tilde C}}}

\def\tq{{{\tilde q}}}

\def\lam{{\lambda}}
\def\blam{{\overline{\lambda}}}
\def\bpsi{{\overline{\psi}}}
\def\bchi{{\overline{\chi}}}
\def\barz{{\overline{z}}}
\def\barv{{\overline{v}}}
\def\barw{{\overline{w}}}

\def\bare{{\overline{e}}}

\def\p{{\partial}}
\def\delbar{{\overline {\partial}}}
\def\Delbar{{\overline {D}}}

\def\wdg{{\wedge}}
\def\vert{{|}}

\def\tg{{\tilde g}}
\def\tP{{\tilde P}}

\def\F{{\tilde F}}

\title{The algebra of Wilson-'t Hooft operators}

\author{Anton Kapustin, Natalia Saulina\\{\small \it California Institute of Technology, Pasadena, CA 91125,
U.S.A.}}

\begin{document}

\begin{titlepage}

\maketitle

\begin{abstract}
We study the Operator Product Expansion of Wilson-'t Hooft operators
in a twisted $\cN=4$ super-Yang-Mills theory with gauge group $G$.
The Montonen-Olive duality puts strong constraints on the OPE and in
the case $G=SU(2)$ completely determines it. From the mathematical
point of view, the Montonen-Olive duality predicts the $L^2$
Dolbeault cohomology of certain equivariant vector bundles on
Schubert cells in the affine Grassmannian. We verify some of these
predictions. We also make some general observations about higher
categories and defects in Topological Field Theories.

\end{abstract}

\vspace{-7in}
\parbox{\linewidth}
{\small\hfill \shortstack{CALT-mm-nnnn}} \vspace{6in}

\end{titlepage}

\section{Introduction}

An important property of Yang-Mills theory is that it contains
Wilson loop operators labeled by irreducible representations of the
gauge group $G$ \cite{Wilson}. Their product is controlled by the
representation ring of $G$ and therefore determines $G$ uniquely.
The work of Goddard, Nuyts, and Olive \cite{GNO} on magnetic sources
can be reinterpreted \cite{KWH} as saying that Yang-Mills theory
admits another class of loop operators labeled by irreducible
representations of the Langlands-dual group $\LG$. Such operators
are called 't Hooft loop operators. The Montonen-Olive duality
conjecture \cite{MO} states that $\cN=4$ super-Yang-Mills theory
with gauge group $G$ is isomorphic to $\cN=4$ super-Yang-Mills
theory with gauge group $\LG$, and this isomorphism exchanges Wilson
and 't Hooft loop operators. This conjecture therefore predicts that
the product of 't Hooft loop operators is controlled by the
representation ring of $\LG$.

This implication of the Montonen-Olive conjecture has been verified
in \cite{KW} for suitably supersymmetrized versions of 't Hooft
loops. The idea is to twist $\cN=4$ SYM theory into a 4d Topological
Field Theory (TFT), so that either Wilson or 't Hooft loop operators
become topological observables. One can show then that the product
of loop operators is independent of the distance between them, and
in fact loop operators form a commutative ring. In the case of
Wilson loop operators, it is straightforward to show that this ring
is the representation ring of $G$. In the case of 't Hooft loop
operators, it has effectively been argued in \cite{KW} that the ring
is the $K^0$-group of the category of equivariant perverse sheaves
on the affine Grassmannian $Gr_G$. It has been shown by Lusztig
\cite{Lusztig} that this ring is the representation ring of $\LG$; a
categorification of this statement, known as the geometric Satake
correspondence, has been proved in \cite{Ginz,MV1,MV2}. As explained
in \cite{KW}, the geometric Satake correspondence can also be
interpreted in physical terms, by replacing loop operators with line
operators.

Yang-Mills theory also admits mixed Wilson-'t Hooft loop operators.
As explained in \cite{KWH}, they are labeled by elements of the set
$$
\dL(G)/\cW=(\Lambda_w(G)\oplus \Lambda_w(\LG))/\cW,
$$
where $\Lambda_w(G)$ is the weight lattice of $G$ and $\cW$ is the
Weyl group (which is the same for $G$ and $\LG$). It is natural to
ask what controls the product of such more general operators. The
answer must somehow unify the representation theory of $G$ and
$\LG$. In this paper we partially answer this question. A natural
framework for it is the holomorphic-topological twisted version of
the $\cN=4$ SYM theory described in \cite{htft}, since it admits
Wilson-'t Hooft loop operators labeled by arbitrary elements of
$\dL/\cW$.\footnote{In the topological field theory described in
\cite{KW}, depending on the choice of a BRST operator, either Wilson
or 't Hooft loop operators may exist, but not both at the same time.
In what follows we will refer to this TFT as the GL-twisted theory,
where GL stands for ``geometric Langlands''.} As explained in
\cite{htft}, Wilson-'t Hooft loop operators in the twisted theory
form a commutative ring, and this ring is abstractly isomorphic to
the Weyl-invariant part of the group algebra $\dL(G)$. But this does
not completely determine the operator product, since we do not yet
know which element of the group algebra corresponds to a particular
element of the set $\dL(G)/\cW$ labeling Wilson-'t Hooft loop
operators.

In this paper we determine the answer for $G=PSU(2)$ and $G=SU(2)$
assuming S-duality, and then verify the prediction in a special case
by a direct gauge-theory computation at weak coupling. We also
outline a procedure for computing the product of Wilson-t' Hooft
loop operators for arbitrary $G$. The procedure is very similar to
that for 't Hooft operators in \cite{KW}. As in \cite{KW}, an
important role is played by the fact that loop operators can be
promoted to line operators, i.e. ``open'' analogs of loop operators.
While loop operators form a commutative ring, line operators form a
monoidal category (i.e. an additive category with a ``tensor
product''). We argue below that the ring of loop operators can be
thought of as the $K^0$-group of the category of line operators. The
Montonen-Olive duality predicts that these categories for gauge
groups $G$ and $\LG$ are equivalent. In some sense, this can be
viewed as the classical limit of the geometric Satake
correspondence, but $G$ and $\LG$ enter more symmetrically. As
discussed in the concluding section, this conjecture, when
interpreted in mathematical terms, has previously appeared in
\cite{BFM}.

\section{A brief review of the Hitchin moduli space}

In this preliminary section we review some basic facts about the
moduli space of Hitchin equations $\MH(G,C)$ and the sigma-model
with target $\MH(G,C)$. The reader familiar with this material may
skip this section. A more detailed discussion may be found in
\cite{KW}.

Given a gauge group $G$, let us consider a principal $G$-bundle $E$
over a Riemann surface $C$, a connection $A$ on $E$, and a 1-form
$\phi$ with values in $\ad(E)$. The Hitchin equations are
$$F-i\, \phi\wdg \phi=0,\quad D\phi=0,\quad D\star \phi=0,$$
where $D=d+iA$ is the covariant differential, $F=-iD^2$ is the
curvature of $A$, and $\star$ is the Hodge star operator. The space
of solutions of this equations modulo gauge transformations is known
as the Hitchin moduli space and will be denoted $\MH(G,C)$ or simply
$\MH$ (we suppress $E$ from the notation, because we regard
$\MH(G,C)$ as a disconnected sum of components corresponding to all
possible topological types of $E$).

A crucial fact for us is that ${\cal M}_H$ is a hyperk\"ahler
manifold. In particular, it has three complex structures $I,J,K$
satisfying $IJ=K$. One way to describe these complex structures
explicitly is to specify holomorphic coordinates on $\MH$. For a
local complex coordinate $z$ on $C$ we write
$$A=A_z dz+A_{\bz}d{\bz},\quad \phi=\phi_z dz +\phi_{\bz}d{\bz}.$$
For the complex structure $I$ the holomorphic coordinates are
$A_{\bz}$ and $\phi_z.$ For the complex structure $J$ the
holomorphic coordinates are $A_{\bz}+i\phi_{\bz}$ and $A_z+i\phi_z.$
Finally, the complex structure $K$ is defined by the quaternion
relation $K=IJ.$ In the present paper we mostly work with complex
structure $I$ and use notation ${\cal M}_{Higgs}(G,C)$ for ${\cal
M}_H(G,C)$ with this choice of complex structure. The reason for
this notation is that $\MH$ equipped with the complex structure $I$
is naturally identified with the moduli space of Higgs bundles, i.e.
pairs $(\cE,\varphi)$, where $\cE$ is a holomorphic $G_\CC$ bundle,
and $\varphi$ is a holomorphic section of $K_C\otimes \ad(\cE)$.
This identification maps the triple $(E,A,\phi)$ to the holomorphic
$G_\CC$-bundle defined by the $(0,1)$ part of $D$ and the
holomorphic Higgs field $\varphi=\phi^{1,0}$. Note that the subset
of $\cM_{Higgs}(G,C)$ given by $\varphi=0$ is the moduli space of
stable holomorphic $G_\CC$ bundles, which we will denote $\cM(G,C)$.

In the complex structure $J$ the Hitchin moduli space can be
identified with the moduli space of flat $G_\CC$ connections on $C$;
this moduli space was denoted $\cM_{flat}(G,C)$ in \cite{KW}. But
this identification will not play a role in this paper.

Consider now the supersymmetric sigma-model with target $\MH$. Since
$\MH$ is hyperk\"ahler, such a sigma-model has $\cN=(4,4)$
supersymmetry. One may twist this sigma-model into a topological
field theory by picking a pair of complex structures $(J_+,J_-)$ on
$\MH(G,C)$. For $J_+=J_-$ one gets a B-model, while for $J_+=-J_-$
one gets an A-model. In this paper we will be mostly interested in
the special case $J_+=J_-=I$, i.e. the B-model in complex structure
$I$.

Given a topological twist of the sigma-model, one can consider the
corresponding category of topological branes. This is a category of
boundary conditions for the sigma-model on a worldsheet of the form
$\RR\times {\rm I}$ where ${\rm I}$ is the unit interval. The
boundary conditions are required to be invariant with respect to the
BRST operator of the twisted model. Equivalently, one may say that
the boundary conditions are required to preserve one complex
supercharge (in the untwisted theory). But since the untwisted model
has $(4,4)$ supersymmetry, there also exist branes which preserve
two complex supercharges. Such branes are compatible with more than
one topological twist. In this paper we will encounter
$(B,B,B)$-branes, which are B-branes in complex structures $I,J,K,$
as well as $(B,A,A)$-branes which are of  B-type in complex
structure $I$ and of $A-$type in the other two complex structures.

\section{Holomorphic-topological twist of $\cN=4$ SYM}\label{twist}

Let us recall how one can twist $\cN=4$ gauge theory on
$\Sigma\times C$ into a holomorphic-topological theory \cite{htft}
which upon reduction gives the B-model on $\Sigma$ with target
$\cM_{Higgs}(G,C)$. It is convenient to treat $\cN=4$ SYM as $\cN=2$
SYM with a hypermultiplet in the adjoint representation. The theory
has $SU(2)_R\times U(1)_N\times U(1)_B$ symmetry. The holonomy group
is $U(1)_C\times U(1)_\Sigma$. One twists $U(1)_C$ action by a
suitable linear combination of $U(1)_R\subset SU(2)_R$ and $U(1)_B$,
and twists $U(1)_\Sigma$ by $U(1)_N$.

The resulting field theory has the following bosonic fields: the
gauge field $A$, the adjoint Higgs field $\vphi=\Phi_w dw\in
K_\Sigma\ot \ad(E)$, the adjoint Higgs field $q=q_\bz d\bz\in
\bK_C\ot \ad(E),$ and the adjoint Higgs field $\tq\in \ad(E)$. Here
$K_\Sigma$ and $K_C$ are the pull-backs of the canonical line
bundles of $\Sigma$ and $C$ to $\Sigma\times C$. We also define
$\Phi_\bw=\Phi_w^\dag$ and $q_z=q_\bz^\dag$.

The fermionic fields are the ``gauginos''
$\lambda_w,\blambda_\bw,\lambda_z, \blambda_z, \lambda_{\bz w},
\blambda_{\bz \bw}, \lambda_{w\bw},\blambda_{w\bw}$ and the
``quarks'' $\psi_\bw, \bchi_w, \psi_\bz, \bchi_\bz, \chi_{z\bw},
\bpsi_{zw}, \chi_{z\bz}, \bpsi_{z\bz}.$ The fermions are all in the
adjoint representation.

The field content depends on complex structures of $C$ and $\Sigma$.
The dependence on the complex structure on $C$ is inescapable, but
the dependence on the complex structure on $\Sigma$ is merely an
artifact of our way of presentation. It is possible to combine
fields with holomorphic and anti-holomorphic indices into
form-valued fields on $\Sigma$ so that the dependence on the complex
structure on $\Sigma$ is eliminated \cite{htft}.

In order to specify the theory completely, one has to pick a BRST
operator. The twisted theory has two BRST operators $Q_\ell$ and
$Q_r$ which square to zero and anticommute, so the most general BRST
operator is
$$
Q=u Q_\ell +v Q_r,
$$
where $u,v$ are homogeneous coordinates on $\PP^1$. It is often
convenient to work with an affine coordinate $t=v/u$ taking values
in $\CC\cup \{\infty\}$. To get a theory which is topological on
$\Sigma$ and holomorphic on $C$, one needs to assume that $u$ and
$v$ are both nonzero, i.e. $t\neq 0,\infty$ \cite{htft}.\footnote{If
$t=0$ or $t=\infty$, the twisted theory is holomorphic on both $C$
and $\Sigma$. Such a theory does not admit line operators which we
are interested in.} The precise choice of $t$ then does not matter
\cite{htft}; we let $t=i$ from now on.

The action of the twisted theory can be written as a sum of a
BRST-exact piece and a piece which is independent of the gauge
coupling $e^2$ and the $\theta$-parameter (after a rescaling of
fermions). Therefore semiclassical computations in the twisted
theory are exact \cite{htft}. We will use this important fact
throughout the rest of the paper.

The path-integral of the twisted theory localizes on $Q$-invariant
field configurations. The conditions of $Q$-invariance imply, among
other things, that the complex connection $\cA=A+i\vphi+i\vphi^\dag$
has a curvature $\cF$ whose only nonzero components are along
$\Sigma$. In the limit when the volume of $C$ goes to zero, the
equations simplify and imply the Hitchin equations for $A_z$ and
$q_z$
$$
F_{z\bz}-i[q_z,q_\bz]=0,\quad D_\bz q_z=0
$$
as well as
$$
D_\bz \tq^\dag=0,
$$
which implies that $\tq$ is generically zero. Thus in this limit the
field theory reduces to a sigma-model with target
$\cM_{Higgs}(G,C)$. There are further equations which say that this
sigma-model is a B-model in the natural complex structure (the one
which we denote $I$).

The Montonen-Olive duality, as usually defined, maps $G$ to $\LG$
and maps \cite{KW,htft} the BRST operator at $t=i$ to another BRST
operator with
$$
\Lt=\frac{|\tau|}{\tau}t.
$$
But since the phase of $t$ can be changed by an automorphism of the
theory (an R-symmetry transformation), one can redefine the
Montonen-Olive duality so that it leaves $t$ invariant. We adopt
this definition of Montonen-Olive duality from now on.

In this paper we mostly focus on the case when $\Sigma$ has a flat
metric. Then the twist along $\Sigma$ is a trivial operation, and
the theory can be regarded as twisted only along $C$. In the limit
$\vol(C)\ra 0$ it becomes equivalent to an untwisted supersymmetric
sigma-model with target $\MH(G,C)$. Since $\MH$ is hyperk\"ahler,
this sigma-model has $\cN=(4,4)$ supersymmetry, i.e. it has two
left-moving and two right-moving complex supercharges, as well as
their complex conjugates. The BRST operator defined above is a
particular linear combination of these supercharges. The BRST
operator of the GL twisted theory considered in \cite{KW} is another
such linear combination (depending on a single complex parameter
$t$). Both kinds of BRST operators can be included into a more
general three-parameter family of BRST operators \cite{KW}.

\section{Wilson-'t Hooft operators in the twisted theory}\label{wh}

\subsection{Definition}

In any gauge theory one can define various loop operators: Wilson,
't Hooft, and Wilson-'t Hooft. The Wilson loop operator in
representation $R$ is usually defined as
$$
W_R(\gamma)=\Tr_R\, P\exp i\int_\gamma A
$$
where $\gamma$ is a closed curve. Instead of labeling the operator
by an irreducible representation, one can label it by the orbit of
its highest weight under the Weyl group. The 't Hooft loop operator
is a disorder operator defined by the requirement that near a curve
$\gamma$ the gauge field has a singularity of a Dirac-monopole kind.
Such singularities are labeled by conjugacy classes of homomorphisms
from $U(1)$ to $G$, which is equivalent to saying that they are
labeled by orbits of the Weyl group in the coweight lattice
$\Lambda_{cw}$ of $G$. More generally, Wilson-'t Hooft operators are
labeled by Weyl orbits in the product $\Lambda_w(G)\times
\Lambda_{cw}(G)$ \cite{KWH}.

In the $\cN=4$ SYM theory there are more possibilities for loop
operators, since one can construct them not only from gauge fields,
but also from other fields. By imposing natural symmetry
requirements (namely, the geometric symmetries and supersymmetry),
one can cut down on the number of possibilities.

In the twisted $\cN=4$ theory we have to require that loop operators
be BRST-invariant. For $t=i$, we see that none of the components of
$A$ are BRST-invariant. But we also see that $\cA_w=A_w+i\Phi_w$ and
$\cA_\bw=A_\bw+i\Phi_\bw$ are BRST-invariant. Hence if $\gamma$ is a
closed curve on $\Sigma$ and $p$ is a point on $C$, the Wilson
operator
$$
W_R(\gamma,p)=\Tr_R\, P\exp i \int_{\gamma\times p}\cA
$$
is BRST-invariant.

By MO duality, there should also be BRST-invariant 't Hooft
operators at $t=i$.\footnote{This is unlike the GL twisted theory,
where for $t=i$ only Wilson operators are BRST-invariant.} Indeed,
if $\gamma$ is given by the equation $x^1=\Re w=0$ and we require
the gauge field to have a Dirac-like singularity in the
$x^1,x^2,x^3$ plane:
\begin{equation}\label{HopF}
F\sim \star_3 d\left(\frac{\mu}{2r}\right)
\end{equation}
for some $\mu\in\frg$, then the condition of $Q$-invariance requires
$\Phi_w$ to be singular as well:
\begin{equation}\label{Hopphi}
\Phi_w\sim \frac{\mu}{2r}.
\end{equation}
It is a plausible guess that such a disorder operator is mapped to
the Wilson operator by the MO duality.

Finally, we may consider more general Wilson-'t Hooft loop operators
which source both electric and magnetic fields. Roughly speaking,
they are products of Wilson and 't Hooft operators. To define a WH
loop operator more precisely, let it be localized at $x^{1,2,3}=0$.
Then we require the components of the curvature in the $123$ plane
to have a singularity as in (\ref{HopF}), the real part of $\Phi_w$
to have a singularity as in (\ref{Hopphi}), and insert into the
path-integral a factor
$$
\Tr_{R}\, P\exp i \int_{\gamma\times p}\cA
$$
where $R$ is an irreducible representation of the stabilizer
subgroup $G_\mu\subset G$ of $\mu$. This definition makes sense
because in the infinitesimal neighborhood of $\gamma$ the component
of $\cA$ tangent to $\gamma$ must lie in the centralizer subalgebra
$\frg_\mu\subset\frg$ of $\mu$ \cite{KWH}. One may describe $R$ by
specifying its highest weight $\nu$, which is defined up to an
action of the subgroup of the Weyl group which preserves $\mu$. The
net result is that the WH operator is labeled by a pair
$(\mu,\nu)\in \Lambda_{cw}(G)\times \Lambda_w(G)$ defined up to the
action of the Weyl group $\cW$. We will denote the abelian group
$\Lambda_{cw}(G)\times \Lambda_w(G)$ by $\dL(G)=\dL(\LG)$. The WH
operator labeled by the Weyl-equivalence class of $(\mu,\nu)$ will
be denoted $WT_{\mu,\nu}(\gamma,p)$.

There is a natural action of the S-duality group on $\dL(G)$. It is
a natural conjecture that this is how the S-duality group acts on
the corresponding WH operators. One of the goals of this paper is to
test this conjecture.

Note that all our loop operators are localized at points on $C$. If
we take the volume of $\Sigma$ to be small compared to that of $C$,
then the twisted theory reduces to an effective 2d field theory on
$C$, and in this effective 2d field theory our loop operators behave
in all ways like local operators. There are no BRST-invariant
operators which are localized on loops in $C$.

\subsection{Basic properties}

As explained in \cite{htft}, in the twisted theory all correlators
depend holomorphically on coordinates on $C$ and are invariant under
arbitrary diffeomorphisms of $\Sigma$. This puts strong constraints
on the correlators of WH loop operators. We will be mostly
interested in the Operator Product Expansion of WH loop operators.
That is, we will assume that $\Sigma$ is flat, pick a pair of points
$p,p'\in C$ and a pair of straight lines $\gamma$ and $\gamma'$ on
$\Sigma$ and consider a pair of WH operators localized on
$\gamma\times p$ and $\gamma'\times p'$. So far, we have assumed
that the curve on which the WH operator is localized is closed; if
we want to maintain this, we may assume that $\Sigma$ locally looks
like a cylinder with a flat metric; since the theory is
diffeomorphism-invariant along $\Sigma$, the only thing that matters
is that both $\gamma$ and $\gamma'$ are closed and isotopic to each
other. One may also consider WH operators localized on lines rather
than closed curves; we will return to this possibility later.

Consider now a correlator involving these WH loop operators. If
$\gamma$ and $\gamma'$ do not have common points, then there is no
singularity as one takes the limit where $p$ coincides with $p'$. If
$z$ is a local complex coordinate on $C$ centered at $p$, then the
correlator is a holomorphic function of $z(p')$ in the neighborhood
of zero. By continuity, this implies that even when $\gamma$ and
$\gamma'$ coincide, the correlator is a holomorphic function of $z$.
Therefore the Operator Product of any two WH operators is
nonsingular. More generally, this conclusion holds for any two
BRST-invariant loop operators in the twisted theory which are
localized on $C$.

Given this result, we can define a commutative algebra of loop
operators, simply by taking the coincidence limit. For Wilson and 't
Hooft loop operators this result can be more easily obtained using
the GL twisted theory of \cite{KW}, but here we see that it holds
for general loop operators in the holomorphic-topological twisted
theory.

At this stage it is natural to ask whether the subspace spanned by
WH operators is closed with respect to the operator product. More
optimistically, one could hope that WH operators form a basis in the
space of loop operators in the twisted theory, and therefore the
vector space spanned by them is automatically closed with respect to
the operator product. We will argue below that both statements are
true, {\it if only closed loops are considered}.

\subsection{Line versus loop operators}

As emphasized in \cite{KW}, one may also consider analogs of Wilson
and 't Hooft operators localized on open curves instead of loops.
The endpoints of a curve must lie on the boundaries of the
four-manifold. Such ``operators'' are called line operators in
\cite{KW}. We put the word ``operators'' in quotes because they do
not act on the Hilbert space of the theory; rather, they alter the
definition of the Hilbert space of the theory.

To be concrete, suppose $\Sigma=\RR\times X_1$, where $X_1$ is
either $S^1$ or an interval $I$. We regard $\RR$ as the time
direction. Consider a Wilson line operator $W_R(\gamma,p)$, where
$\gamma\subset \Sigma$ has the form $\RR\times q$ for some $q\in
X_1$. Insertion of such a Wilson line operator means that the
Hilbert space of the gauge theory has to be modified: instead of
gauge-invariant wave-functions on the space of fields on $X_1\times
C$, one has to consider gauge-invariant elements of the tensor
product of the space of all wave-functions and the representation
space of $R$. Similarly, when we insert an open 't Hooft operator,
we have to change the class of fields on which the wave-functions are
defined.

While loop operators form a commutative algebra, line operators form
a category. A morphism between line operators $\sA$ and $\sB$ is a
local BRST-invariant operator inserted at a junction of $\sA$ and
$\sB$. Composition of morphisms is defined in an obvious way. There
is also an obvious structure of a complex vector space on the space
of morphisms and an obvious way to define a sum of line operators.
Thus line operators form an additive $\CC$-linear category.

The distinction between line and loop operators has played some role
in \cite{KW} and it is even more important in the context of the
holomorphic-topological theory, as we will see below.

It is often convenient to relax the condition that local operators
inserted at the junction of two line operators be BRST-invariant,
and define the space of morphisms to be the space of all local
operators. This space is graded by the ghost number and is acted
upon by the BRST-differential. Thus the set of morphisms between any
two line operators has the structure of a complex of vector spaces,
and composition of morphisms is compatible with the differentials.
That is, line operators form a differential graded category
(DG-category). This viewpoint is convenient for keeping track of the
dependence of various correlators on parameters, such as the
insertion point on $C$ (see below).

There is one more important operation for line operators in the
twisted theory: an associative tensor product. In other words, the
category of line operators is a monoidal category. The product is
defined by taking two line operators ``side-by-side'' on $\Sigma$
and ``fusing'' them together. The product of line operators need not
be commutative, in general. But for Wilson-'t Hooft line operators
it is commutative because of a discrete symmetry: parity reversal.
Indeed, consider the twisted gauge theory on $\RR\times\RR\times C$,
where we regard the first copy of $\RR$ as time and the second one
as space. It is easy to check that spatial reflection $x\ra -x$ is a
symmetry of the theory.\footnote{This is particularly obvious from a
2d viewpoint, as any B-model is parity-invariant.} Furthermore,
Wilson-'t Hooft line operators are invariant under this symmetry.
Therefore, we can change the order of WH line operators on the
spatial line by a symmetry transformation.

\subsection{Remarks on TFT in arbitrary dimension}

A similar discussion applies to the GL twisted theory considered in
\cite{KW}, and in fact to any topological field theory in any number
of dimensions. That is, in any TFT line operators form a monoidal
$\CC$-linear additive category.

In the case of a TFT in dimension $d>3$ the fusion product is
necessarily symmetric, because there is no diffeomorphism-invariant
way to order line operators. In dimension $d=3$ there may be
nontrivial braiding, so in general the category of line operators is
braided rather than symmetric. A well-known example is the
Chern-Simons theory \cite{WittenCS}, where the category of Wilson
line operators is equivalent to the category of representations of a
quantum group. In dimension $d=2$ the monoidal structure need not be
either symmetric or braided, in general.

In this paper we are dealing with a holomorphic-topological field
theory rather than a TFT, and the ``topological'' part of the
manifold is two-dimensional. From the abstract viewpoint the
situation is very much like in a 2d TFT, because every line operator
in the twisted gauge theory on $\Sigma\times C$ can be regarded as a
line operator in the B-model on $\Sigma$ with target
$\cM_{Higgs}(G,C)$. But the converse is not necessarily true,
because line operators in gauge theory are local on $C$, while line
operators in the B-model on $\Sigma$ are not subject to this
constraint. (Below we will construct a large class of examples of line
operators in the B-model which do not lift to ordinary line
operators in the gauge theory.) To enforce locality, one has to keep
track of the dependence of all correlators on the insertion point
$p\in C$ of the line operator. To put it differently, if we denote
by $\sV(q,p)$ the Hilbert space of the twisted theory on $\RR\times
X_1 \times C$ with an insertion of a line operator at $q\times p\in
X_1\times C$, then for fixed $q$ this family of vector spaces can be
thought of as a holomorphic vector bundle $\sV_q$ over $C$.
Similarly, spaces of morphisms between different line operators can
be thought of as holomorphic vector bundles over $C$.

To make precise the idea of a ``holomorphically varying space of
morphisms'', it is very convenient to take the viewpoint that the
space of morphisms is a differential graded vector space, i.e. a
complex. Let $\sU(p)$ be the vector space of all (not necessarily
BRST-invariant) local operators inserted at the junction of two line
operators $\sA$ and $\sB$, both located at a point $p\in C$. The
space $\sU(p)$ is graded by the ghost number and carries the
BRST-differential $Q$. The complexes $\sU(p)$ fit into a complex of
smooth vector bundles $\sU$ on $C$. Let us tensor this complex of
vector bundles with the Dolbeault complex of $C$. The resulting
space of sections is acted upon by both $Q$ and $\bpartial$ and
carries all the information about the dependence of morphisms on
$p$. ``Holomorphic dependence'' means simply that $\bpartial$ is
$Q$-exact, and therefore acts trivially on the cohomology of $Q$.

We can put our discussion of line operators in a more general
perspective by noting that $n$-dimensional TFTs form a $n$-category.
1-Morphisms in this $n$-category are codimension-1 walls separating
a pair of TFTs. We will call codimension-1 walls 1-walls, for short.
1-walls themselves form an $n-1$ category: 2-morphisms are
codimension-2 walls which separate different 1-walls between the
same pair of TFTs. And so on.

If we consider all 1-walls between a pair of identical TFTs, they
can be ''fused'' together. This gives a kind of monoidal structure
on an $n-1$ category of 1-walls. In this $n-1$-category there is a
unit object: the ``trivial 1-wall'' which is equivalent to no wall
at all. 2-walls living on the trivial 1-wall form a monoidal $n-2$
category with a unit, and so on. Thus line operators considered
above belong to a rather special variety: they live on a trivial
$n-2$ wall which lives on a trivial $n-3$-wall, etc. For example, in
the GL twisted theory at $t=i$ Wilson line operators form a category
which is equivalent to the category of finite-dimensional
representations of $G$. Gukov and Witten also considered nontrivial
2-walls in this theory and line operators living on such 2-walls
\cite{GW}.

Boundary conditions for an $n$-dimensional TFT also fit into this
general scheme: they are 1-morphisms between a given TFT and an
``empty'' TFT. For this reason they form an $n-1$ category (which is
not monoidal, in general). A special case of this is the well-known
fact that D-branes in a 2d TFT form a category.

In connection with possible 2-dimensional generalizations of the
Geometric Langlands Duality, it would be interesting to understand
the 3-category of boundary conditions for the GL twisted $\cN=4$
SYM, as well as the monoidal 3-category of 1-walls in the same
theory. The latter acts on the former. These 3-categories appear to
be suitable 2d generalizations of the derived category of
$\cM_{flat}(G,C)$ and the representation category of $G$,
respectively.

\subsection{Deformations of line operators}

In the case of the GL twisted theory at $t=i$ the product of two
parallel Wilson loop operators $W_{R_1}$ and $W_{R_2}$ is a Wilson
loop operator $W_{R_1\otimes R_2}$. This means that Wilson loop
operators form a closed algebra, which happens to be commutative and
associative. Wilson loop operators corresponding to irreducible
representations of $G$ form a basis in this algebra. A similar
statement holds for Wilson line operators: the subcategory of Wilson
line operators is closed with respect to the monoidal structure,
i.e. it is a symmetric monoidal category, and any Wilson line
operator is isomorphic to a direct sum of Wilson line operators
corresponding to irreducible representations of $G$. By S-duality,
similar statements hold for 't Hooft operators in the GL twisted
theory (for $t=1$).

At $t=i$ any line operator in the GL-twisted theory is isomorphic to
a Wilson line operator for some $R$ (which can be reducible). One
way to see it is to first classify line operators with the right
bosonic symmetries in the untwisted theory (this has been done in
\cite{KWH}) and then impose the condition of BRST-invariance. A
similar statement holds for 't Hooft operators at $t=1$.

One consequence of this is that there are no infinitesimal
deformations of Wilson line operators in the GL-twisted theory. This
can also be checked directly. From the mathematical viewpoint,
infinitesimal deformations of a line operator $\sA$ are classified
by degree-1 cohomology of the complex $\Hom(\sA,\sA)$. One can check
that this cohomology is trivial by considering BRST-invariant local
operators which can be inserted at a point of the Wilson line $\sA$.

For line operators in the holomorphic-topological twisted theory the
situation is more complicated. The difficulty is that twisting
breaks $SO(3)$ rotational symmetry used in \cite{KWH} down to
$U(1)$. A generic Wilson-'t Hooft operators (i.e. not purely
electric or purely magnetic) also preserves only rotation symmetry
in the $z$-plane (which is present when $C\simeq \CC$).

The simplest question one can ask in this regard is whether there
are infinitesimal deformations of a Wilson-'t Hooft line operator.
One obvious deformation arises from varying the insertion point on
$C$. For a Wilson line $W_R(p)$, is easy to exhibit the degree-1
endomorphism corresponding to such a deformation. It is a fermionic
field
$$
\Gamma_z=\lambda_z+\blambda_z.
$$
It is BRST-invariant and can be inserted into a Wilson line in any
representation $R$. The corresponding infinitesimal deformation of
$W_R(p)$ is obtained as follows. First, we apply the descent
procedure to $\Gamma_z$, i.e. look for a boson $\Delta_z$ such that
$$
\cD_\Sigma \Gamma_z=\delta \Delta_z.
$$
Note the covariant differential on the left-hand side. Usually,
descent is applied to gauge-invariant operators, in which case one
uses ordinary de Rham differential. In our case, the operator
becomes gauge-invariant only after insertion into a Wilson line, and
this requires replacing ordinary differential with the covariant
one. The descent equation is solved by
$$
\Delta_z=\cF_{zw} dw+\cF_{z\bw} d\bw.
$$
The deformed Wilson operator is
$$
\Tr_R P\exp\left(i\int \cA+\Delta_z \eps^z\right)
$$
where $\eps^z$ is an infinitesimal parameter. It is easy to see that
this is the same as a Wilson operator evaluated at a nearby point,
shifted from $p$ by a vector $\eps^z\partial_z.$

Similarly, given any two line operators and a degree-$1$ morphism
between them, one can construct their ``bound state'', which is a
deformation of the direct sum of the two line operators. In
homological algebra, this is known as the mapping cone construction.
In section 6.1 we will see some examples of the mapping cone construction
with less obvious deformations of Wilson-'t Hooft line operators
which do not correspond to changing the insertion point on $C$.

\subsection{Line operators and K-theory}

The existence of nontrivial deformations suggests that the category
of Wilson-'t Hooft line operators may not be closed with respect to
the tensor product. But we will argue below that the space of
Wilson-'t Hooft {\it loop} operators is closed with respect to the
product. Therefore it is important to understand the relationship
between loop and line operators. We would like to argue here that
loop operators should be thought of as elements of the $K^0$-group
of the category of line operators. The closure of the space of
Wilson-'t Hooft loop operators under operator product suggests that
these operators form a basis for the $K^0$-group of the category of
line operators, but we will not try to prove this here.

First, let us recall the definition of the $K^0$-group of a
DG-algebra $\cA$. A finitely-generated projective DG-module over
$\cA$ is any DG-module which can be obtained from free DG-modules of
finite rank using the following three operations: shift of grading,
cone, and taking a direct summand. Consider a free abelian group
generated by the isomorphism classes of finitely-generated
projective DG-modules and quotient it by the relations
$$
M\sim (-1)^n M[n]
$$
for any integer $n$, and
$$
M_1\oplus M_2\sim M
$$
for any exact sequence of DG-modules
$$
0\ra M_1\ra M\ra M_2\ra 0.
$$
This quotient group is $K^0(\cA)$.

The definition of the $K^0$-group of a small DG-category $\frA$ is
similar.\footnote{A small category is a category whose objects are
members of a set rather than a class. We sincerely hope that line
operators in a twisted gauge theory form a set.} The idea is to
think about a category as an ``algebra with several objects''. A
DG-module $\frM$ over a small DG-category $\frA$ is a DG-functor
from $\frA$ to the DG-category of complexes of vector spaces. In
more detail, it is a collection of DG-modules $\frM(\sA)$ over the
DG-algebras $\Hom_\frA(\sA,\sA)$ for all $\sA\in Ob(\frA)$ and
DG-morphisms from the complex $\Hom_\frA(\sA,\sB)$ to the complex
$\Hom(\frM(\sA),\frM(\sB))$ for any $\sA,\sB\in Ob(\frA)$. These
data should satisfy some fairly obvious compatibility conditions.

The analog of a free rank-1 DG-module is a presentable DG-module
$\frM_\sB$ corresponding to an object $\sB$ of $\frA$. Given any
$\sB\in Ob(\frA)$, we let $\frM_\sB(\sA)=\Hom_\frA(\sB,\sA)$. It has
an obvious DG-module structure over the DG-algebra
$\Hom_\frA(\sA,\sA)$. A finitely-generated projective DG-module over
$\frA$ is a DG-module which is obtained from presentable modules by
the operations of shift, cone, and taking a direct summand. To get
the $K^0$-group of $\frA$, we consider the free abelian group
generated by isomorphism classes of finitely-generated projective
DG-modules and quotient it by the relations coming from shift of
grading and short exact sequences of DG-modules.

Now let $\frA$ be the DG-category of line operators. A presentable
DG-module corresponding to a line operator $\sB$ is a module
$\frM_B$ such that $\frM_\sB(\sA)$ is the space of local operators
which can be inserted at the joining point of line operators $\sB$
and $\sA$. There is a special line operator: the Wilson line
corresponding to the trivial representation of $G$. It is a unit
object with respect to the monoidal structure on $\frA$. The space
of local operators which can be inserted at such a trivial line
operator is the same as the space of ``bulk'' local operators.

A loop operator is a line operator with no insertions of local
operators and with the endpoints identified. A convenient geometry
to study a loop operator $\sA$ is to take $\Sigma=S^1_\tau\times
S^1_\sigma$, where $S^1_\tau$ is regarded as the compactified
Euclidean time and $S^1_\sigma$ is the compactified spatial
direction. We consider an arbitrary number of insertions of line
operators, one of which is our $\sA$. All line operators are taken
to ``run'' along the $\tau$ direction and are located at fixed
$\sigma$. We also allow arbitrary local insertions at all line
operators except $\sA$. This includes bulk local operator
insertions, which may be regarded as local operators sitting on the
trivial line operator. If all such correlators are unchanged when
one replaces $\sA$ with another loop operator $\sA'$, it is natural
to identify $\sA$ and $\sA'$. We claim that this happens if
$\frA$-modules $\frM_\sA$ and $\frM_{\sA'}$ are in the same $K^0$
class.

To see this, let us reformulate the set-up slightly. First of all,
we can lump all line operators except $\sA$ and all bulk local
operators into a single line operator $\sB$ with a single insertion.
It is easy to see that the Hilbert space of the twisted gauge theory
on $\RR\times S^1_\sigma\times C$ is the homology of the complex
$\Hom_\frA(\sA,\sB)$. Equivalently, we can say that it is the
homology of $\frM_\sA(\sB)$. The local operator inserted into $\sB$
can be thought of as an endomorphism $T$ of the complex
$\frM_\sA(\sB)$, and the correlator is the supertrace of $T$. It is
obvious that shifting the grading of $\frM_\sA$ by $n$  changes the
supertrace by a factor $(-1)^n$. The other equivalence relation has
to do with short exact sequences of $\frA$-modules. If $\frM_\sA$ is
the middle term of a short exact sequence
$$
0\ra \frM_1\ra \frM_\sA\ra \frM_2\ra 0,
$$
then we have a short exact sequence of complexes
$$
0\ra \frM_1(\sB)\ra \frM_\sA(\sB) \ra \frM_2(\sB)\ra 0
$$
and the corresponding long exact sequence in homology. The
endomorphism $T$ of $\frM$ induces an endomorphism $\mathcal T$ of
this long exact sequence, regarded as a complex of vector spaces. We
may assume that both $T$ and $\mathcal T$ are of degree zero, since
otherwise all supertraces vanish for trivial reasons. Now the
statement that the supertrace of $T$ depends only on the $K^0$-class
of $\sA$ is equivalent to the statement that the supertrace of
$\mathcal T$ vanishes. But this is an immediate consequence of
exactness: if $d$ denotes the differential in the long exact
sequence, and $\cR$ denotes the sum of all terms in the long exact
sequence regarded as a graded vector space, then by exactness one
can write
$$
\mathcal T =d\cP+\cP d,
$$
for some linear map $\cP:\cR\ra\cR$ of degree $-1$. The supertrace
of the anticommutator of two odd endomorphisms of a graded vector
space obviously vanishes.

\subsection{Line operators as functors on branes}

We have seen that in the twisted theory line operators form a
monoidal $\CC$-linear category, or, better, a monoidal DG-category.
As in \cite{KW}, it is useful to think of objects of this category
as functors on the category of B-branes on $\cM_{Higgs}(G,C)$. This
makes the monoidal structure more obvious: it is simply given by the
composition of functors.\footnote{Alternatively, one can regard a
B-brane as a 1-morphism between an empty theory and the B-model on
$\cM_{Higgs}(G,C)$, regarded as objects of the 2-category of 2d
TFTs, and one can regard a line operator as a 1-morphism from the
B-model to itself. Then the action of the line operator on the brane
is given by the composition of 1-morphisms.}

It is particularly simple to describe the functor corresponding to a
Wilson line operator $W_R(p)$. It tensors every B-brane on
$\cM_{Higgs}(G,C)$ by a holomorphic vector bundle $R(\cE(p))$, where
$\cE(p)$ is a restriction to $p\in C$ of the universal $G$-bundle
$\cE$ on $\cM_{Higgs}(G,C)\times C$ \cite{KW}.

The functor corresponding to an 't Hooft operator is a Hecke
transformation, as explained in \cite{KW}. Let us remind what a
Hecke transformation is in the case $G=U(N)$. Instead of a principal
$U(N)$-bundle, it is convenient to work with a holomorphic vector
bundle $E$ associated via the tautological $N$-dimensional
representation of $U(N)$. A Hecke transformation of $E_-=E$ at a
point $p\in C$ is another holomorphic vector bundle $E_+$ of the
same rank which is isomorphic to $E_-$ on $C\backslash p$. One can
always choose a basis of holomorphic sections  $f_1,\ldots,f_N$ of
$E_-$ near $p$ so that $E_+$ is locally generated by
$$
s_1=z^{-\mu_1} f_1,\ldots, s_N=z^{-\mu_N} f_N,
$$
where $\mu_1,\ldots,\mu_N$ are integers. The integers
$\mu_1,\ldots,\mu_N$ are well-defined modulo permutation and can be
thought of as a coweight of $U(N)$ modulo the action of the Weyl group.
For fixed $E_-$ and $\mu$, the space of allowed $E_+$ is a
finite-dimensional submanifold $\cC_\mu$ of the infinite-dimensional
affine Grassmannian $GL(N,\CC((z)))/GL(N,\CC[[z]])$, where
$\CC((z))$ is the field of formal Laurent series and $\CC[[z]]$ is
the ring of formal Taylor series. Specifically, $\cC_\mu$ is the
orbit of the matrix
$$
Z_\mu(z)=\diag(z^{-\mu_1},\ldots,z^{-\mu_N})
$$
under the left action of $GL(N,\CC[[z]])$. This describes how 't
Hooft transformations act on structure sheaves of points on
$\cM(G,C)\subset \cM_{Higgs}(G,C)$. One can similarly define the
transformation of a more general point with a nontrivial Higgs
field, see \cite{KW} for details. One can also define how 't
Hooft/Hecke operators act on more general objects of the category of
B-branes, but we will not need this here.

For a general gauge group $G$, the situation is similar. One defines
the affine Grassmannian $Gr_G$ as the quotient $G((z))/G[[z]]$,
where $G((z))$ is the group of $G_\CC$-valued Laurent series and
$G[[z]]$ is the group of $G_\CC$-valued Taylor series. $Gr_G$ is a
union of Schubert cells $\cC_\mu$ labeled by the elements of
$\Lambda_{cw}(G)/\cW$. For a fixed coweight $\mu$ the space of Hecke
transformations of a holomorphic $G$-bundle $E_-$ is the
corresponding Schubert cell $\cC_\mu$.

The functor corresponding to a general Wilson-'t Hooft operator is a
combination of a Hecke transformation and tensoring with a certain
holomorphic vector bundle on $\cC_\mu$. For simplicity, let us only
consider the case when the initial B-brane is a point $E_-$ of
$\cM(G,C)\subset \cM_{Higgs}(G,C)$. Recall that the ``electric''
part  of the Wilson-'t Hooft operator can be described by a
representation $R$ of the group $H$ which is the stabilizer subgroup
of the coweight $\mu$ (under the adjoint representation). Clearly,
the electric degree of freedom will live in some vector bundle over
$\cC_\mu$. This bundle is associated via $R$ to a certain principal
$H$-bundle over $\cC_\mu$.

To determine this bundle, note that over $\cC_\mu$ there is a
principal $G$-bundle whose fiber can be identified with the fiber of
$E_+$ over $z=0$. A formal definition, in the case $G=U(N)$, is as
follows. $\cC_\mu$ can be thought of as the set of equivalence
classes of matrix functions of the form
$$
F(z) Z_\mu(z) G(z), \quad F(z), G(z) \in GL(N,\CC[[z]]), \quad
Z_\mu(z)=\diag(z^{-\mu_1},\ldots, z^{-\mu_N})
$$
under the right action of $GL(N,\CC[[z]])$. Let us now replace
$GL(N,\CC[[z]])$ with its subgroup $GL_0(N,\CC[[z]])$ consisting of
matrix functions which are identity at $z=0$. The set of equivalence
classes of matrices $F(z)Z_\mu(z)G(z)$ under the right
$GL_0(N,\CC[[z]])$ action is clearly a principal $G$-bundle
$\cP_\mu$ over $\cC_\mu$. This $G$-bundle has a reduction to a
principal $P$-bundle $\cQ_\mu$, where $P$ is the parabolic subgroup
whose quotient by the maximal unipotent subgroup is $H_\CC$. (This
reflects the fact that the gauge group is broken down to $H$ near
$z=0$). Explicitly, $\cQ_\mu$ consists of
$GL_0(N,\CC[[z]])$-equivalence classes of matrix functions
$F(z)Z_\mu(z)G(z)$ such that $G(0)\in P$. The group $H$ acts by
right multiplication. The $H_\CC$-bundle we are after is the
quotient of $\cQ_\mu$.

In the next section, we will discuss in detail Wilson-'t Hooft line
operators for $G=PSU(2)$; as a preparation, let us describe the
relevant vector bundles over $\cC_\mu$ in the case when $\mu$ is the
smallest nontrivial coweight. The Schubert cell $\cC_\mu$ in this
case is simply $\PP^1=PSU(2)/U(1)=PSL(2,\CC)/B$, where $B$ is the
Borel subgroup of $G_\CC=PSL(2,\CC)$. The $B$-bundle in question is
simply the tautological bundle $G_\CC\ra G_\CC/B$, and the $H=U(1)$
bundle is the Hopf bundle. The coweight (resp. weight) lattice of
$PSL(2,\CC)$ is isomorphic to the lattice of integers (resp. even
integers). The electric degree of freedom of a Wilson-'t Hooft line
operator with $\mu=1$ and $\nu \in 2 {\bf Z}$ by definition takes
values in the fiber of the line bundle $L$ associated with the Hopf
bundle via a $U(1)$ representation of charge $\nu$. Since the Hopf
bundle is the circle bundle of the line bundle $\cO(-1)$ over
$\PP^1,$ we conclude that $L=\cO(-\nu).$

As a rule, a functor from the derived category of $X$ to itself is
``representable'' by an object of the derived category of $X\times
X$. It is not known whether this is the case for all reasonable
functors, but it is certainly true for functors corresponding to
line operators. To show this, let $\Sigma\simeq \RR^2$, and suppose
for simplicity that the line operator has the shape of a straight
line. Using the ``folding trick'' we can regard the field theory on
$\RR^2$ with an insertion of a straight line operator as a product
of two copies of the same field theory on a half-plane, with a
particular boundary condition. The product of two copies of a
B-model with target $X$ is a B-model with target $X\times X$, and
the boundary condition corresponds to a B-brane on $X\times X$. This
B-brane represents the functor corresponding to the line operator.
For example, in the case of Wilson line operator $W_R(p)$, the
corresponding object is the diagonal of $\cM_{Higgs}(G,C)\times
\cM_{Higgs}(G,C)$ equipped with the holomorphic vector bundle
$R(\cE_p)$.

The ``folding trick'' reduces the study of line operators to the
study of boundary conditions. There is a converse trick which
reduces the study of boundary conditions to the study of line
operators. Consider a B-model on a strip $I\times \RR$ with some
boundary conditions $\alpha$ and $\beta$. We can identify the
$\alpha$ and $\beta$ boundaries and replace $I$ with $S^1$, with an
insertion of a line operator. If we think of $\beta$ as a 1-wall
between our B-model and the empty theory, and about $\alpha$ as the
1-wall between the empty theory and the B-model, then the line
operator is obtained by fusing together these 1-walls to get a
1-wall between the B-model and itself. We will call this the
``gluing trick''.

One application of the ``gluing trick'' is to produce new examples
of line operators from known boundary conditions. Given any two
B-branes on $\cM_{Higgs}(G,C)$ we may produce a line operator in the
B-model with target $\cM_{Higgs}(G,C)$. However, this construction
is not local on $C$ and does not produce new line operators in the
twisted gauge theory. For example, if we start with boundary
conditions for the B-model which can be lifted to the gauge theory,
then ``gluing'' them produces a 3-wall in the gauge theory rather
than a 1-wall. Only upon further compactification on $C$ does one
get a line operator in the 2d TFT.

We have argued above that any correlator involving a loop operator
$\sA$ and any other loop, line, or local operator depends only on
the $K^0$-class of $\sA$. It was assumed that $\Sigma=S^1\times
S^1$. One may ask if the statement remains true if $\Sigma=\RR\times
I$ with suitable boundary conditions. Since the ``gluing trick''
replaces any pair of boundary conditions with a line operator, the
answer appears to be ``yes''. But we have to keep in mind that line
operators produced by the ``gluing trick'' are not local on $C$.
Therefore, to apply the above reasoning we need to work with a
different $K^0$-group: the $K^0$-group of the category of all line
operators in the B-model with target $\cM_{Higgs}(G,C)$.

\subsection{The algebra of loop operators and
S-duality}\label{s:sduality}

We have argued above that loop operators form a commutative algebra.
To identify this algebra, one can use the fact that the gauge theory
becomes abelian in the infrared, if the Higgs field has a generic
expectation value (with all eigenvalues distinct). More precisely,
the gauge group is broken down to a semi-direct product of the
maximal torus $T$ of $G$ and the Weyl group $\cW$. Loop operators in
such a theory are labeled by Weyl-invariant combinations of loop
operators in the abelian gauge theory with gauge group $T$. The
latter are labeled by electric and magnetic charges, i.e. by
elements of $\Hom(T,U(1))=\Lambda_w(G)$ and
$\Hom(U(1),T)=\Lambda_{cw}(G)$. The algebra structure is also
obvious: under Operator Product electric and magnetic charges simply
add up, so the algebra of loop operators is isomorphic to the
Weyl-invariant part of the group algebra of
$\Lambda_w(G)\oplus\Lambda_{cw}(G)=\dL(G)$.

This reasoning may seem suspect, because a vacuum with a particular
expectation value of a Higgs field is not BRST-invariant, and if we
try to integrate over all expectation values, we have to include
vacua where nonabelian gauge symmetry is restored. One can give a
more careful argument as follows. Let us consider again the case
where $M=\Sigma\times C$, and $\Sigma$ has a nonempty boundary. From
the viewpoint of the effective field theory on $\Sigma$, the theory
``abelianizes'' in the limit where the Higgs field $q_z$ is large
and all of its eigenvalues are distinct. The problem is that one has
to integrate over all values of $q_z$, including those where some of
the eigenvalues coincide. To argue that we can perform the
computation locally in the target space $\cM_{Higgs}(G,C)$, recall
that in the B-model the path-integral localizes on constant maps.
Therefore if we impose a boundary condition which keeps $q_z$ away
from the dangerous region, we can be sure that the dangerous regions
of the target space will not contribute. For example, one can take a
boundary condition corresponding to a B-brane which is a generic
fiber of the Hitchin fibration. If $\partial\Sigma$ has several
components, it is sufficient to impose such a boundary condition
only on one component of the boundary.

It is known how the S-duality group acts on the algebra of loop
operators \cite{KWH}. The generator $T,$ which shifts $\tau\ra
\tau+1,$ does not change the magnetic charge $\mu\in\Lambda_{cw}(G)$
and acts on the electric charge $\nu\in\Lambda_w(G)$ by
$$
\nu\ra\nu+\mu.
$$
Here we regard $\mu$ as an element of $\frt^*$ using the
identification of $\frt$ and $\frt^*$, defined by the canonical
metric on $\frt$ (the Killing metric is normalized so that short
coroots have length $\sqrt 2$). The shift of the electric charge is
due to the Witten effect \cite{witteneffect}. The generator $S$
which exchanges $G$ and $\LG$ conjecturally acts by
$$
(\mu, \nu)\ra ( \frR\cdot \mu, \frR\cdot \nu) \begin{pmatrix} 0 &
-1/\sqrt n_\frg \\ \sqrt n_\frg & 0\end{pmatrix}
$$
Here $\frR$ is a certain orthogonal transformation which squares to
an element of the Weyl group \cite{GNO,AKS}. For simply-laced groups
one can define Montonen-Olive duality so that $\frR=1$.

These results, however, do not yet allow us to compute the OPE of
any two given Wilson-'t Hooft operator. To do that, one needs to
know which element of the group algebra of $\dL(G)$ corresponds to
any particular Wilson-'t Hooft operator. Recall that the space of WH
operators has a natural basis labeled by elements of
$$
\dL(G)/\cW
$$
So what we are looking for is a basis for the Weyl-invariant part of
the group algebra of $\dL(G)$ labeled by this set.

The most obvious such basis is obtained simply by taking an element
of $\dL(G)$ in a particular Weyl-equivalence class and averaging it
over the Weyl group. Such basis elements correspond to loop
operators in the abelian gauge theory with particular electric and
magnetic charges.\footnote{Averaging over the Weyl group reflects
the fact that the gauge group is really a semidirect product of the
Weyl group and the maximal torus of $G$.} But this is not the basis
we are looking for. For example, consider a Wilson operator for an
irreducible representation $R_\nu$ with highest weight $\nu$. From
the viewpoint of the effective abelian gauge theory, it is a sum of
Wilson operators with electric charges given by decomposing $R_\nu$
with respect to the maximal torus of $G$. All weights of $R_\nu$
appear in this decomposition, not just the weights which are in the
Weyl-orbit of the highest weight. Similarly, in the phase with the
broken nonabelian gauge symmetry an 't Hooft operator corresponding
to a coweight $\mu$ of $G$ decomposes as a sum over weights of the
representation $\LR_\mu$ of the dual group. The explanation of this
phenomenon is more subtle than for Wilson operators and involves
``monopole bubbling'' \cite{KW}.

In the case $G=PSU(2)$ (or $G=SU(2)$), the desired basis is uniquely
determined by imposing S-duality. To simplify notation, let us
identify the group algebra of $\Lambda_{cw}(PSU(2))\simeq\ZZ$ with
the space of polynomials of $x,x^{-1}$, and the group algebra of
$\Lambda_w(PSU(2))\simeq 2\cdot\ZZ$ with the space of polynomials of
$y^2,y^{-2}$. The Weyl group acts by $x\ra x^{-1}, y\ra y^{-1}$.
Then the algebra of WH loop operators can be identified with the
space of Weyl-invariant polynomials of $x,x^{-1},y^2,y^{-2}$ (for
$G=PSU(2)$) or of $x^2,x^{-2}, y, y^{-1}$ (for $G=SU(2)$). We know
already that the Wilson loop in the representation
with highest weight $n\in\ZZ$ corresponds to the polynomial
$$
WT_{0,n}=y^{n}+y^{n-2}+\ldots+y^{-n}.
$$
Here $n$ is an arbitrary integer if $G=SU(2)$ and an even integer if
$G=PSU(2)$. Similarly, the 't Hooft loop labeled by the coweight
$m\in\ZZ$ corresponds to the polynomial
$$
WT_{m,0}=x^m+x^{m-2}+\ldots+ x^{-m},
$$
where $m\in\ZZ$ if $G=PSU(2)$ and $m\in 2\cdot \ZZ$ if $G=SU(2)$.
This is, of course, compatible with the Montonen-Olive duality,
which acts by
$$
(m,n)\mapsto (n,-m).
$$
Moreover, any pair $(m,n)\in \dL(G)$ can be brought to the form
$(m',0)$ by an S-duality transformation. This determines the
polynomial corresponding to an arbitrary Wilson-'t Hooft operator
for $G=PSU(2)$ or $G=SU(2)$:
$$
WT_{m,n}=x^m y^n+x^{m-2a} y^{n-2b}+x^{m-4a} y^{n-4b}+\ldots + x^{-m}
y^{-n}.
$$
Here the integers $a,b$ are defined by the condition that $m/n=a/b$,
$a$ and $b$ have the same signs as $m$ and $n$, respectively, and
the fraction $a/b$ is reduced.

For higher-rank groups, S-duality is not sufficient to fix the
basis. This is because electric and magnetic charges need not be
linearly dependent for higher-rank gauge groups.

In the next section, we will test some predictions of S-duality for
the gauge group $PSU(2)$ by a direct computation of the OPE of WH
loop operators at weak coupling. The same method could be used to
determine the OPE of WH loop operators for higher-rank groups, but
the computations become very complicated.

\section{OPE at weak coupling}\label{ope}

\subsection{Semiclassical quantization of Wilson-'t Hooft operators}

To compute the OPE of a pair of Wilson-'t Hooft line operators we
will follow the same method as in \cite{KW}. We will quantize the
twisted gauge theory on a manifold with boundaries $C\times I\times
\RR$, with suitable boundary conditions and with two insertions of
Wilson-'t Hooft operators. From the 2d viewpoint, the boundary
conditions correspond to B-branes on $\cM_{Higgs}(G,C)$. The problem
reduces to the supersymmetric quantum mechanics on the space of zero
modes of the gauge theory. In principle, one has to study the limit
where the two operators approach each other, but in the twisted
theory this last step is not necessary, if the line operators are
sitting at the same point on $C$.

As in \cite{KW}, it is convenient to choose the branes so that in
the absence of Wilson-'t Hooft line operators the Hilbert space of
the twisted gauge theory is one-dimensional. One possible choice is
to take the brane $\alpha$ at $y=0$ to be the 0-brane at a point $r$
of $\cM_{Higgs}(G,C)$ with vanishing Higgs field. The brane $\beta$
at $y=1$ will be the trivial line bundle on $\cM_{Higgs}(G,C)$. Both
of these branes are of type $(B,B,B)$. The classical space of vacua
in this case consists of a single point $r$, with no zero modes, so
the Hilbert space is one-dimensional. Alternatively, as in
\cite{KW}, one could take two branes of type $(B,A,A)$ intersecting
at a single point. The former choice is somewhat easier, so we will
stick to it, but in practice there is not much difference between
the two.

Having chosen the boundary conditions, we can assign to any
collection $\sA,\sB,\ldots,$ of WH line operators the graded vector
space $\frV_{\alpha\beta}(\sA,\sB,\ldots)$, or better yet the
corresponding BRST complex. Note that this assignment need not be
invariant with respect to S-duality. This is because the choice of
branes necessarily breaks the S-duality group. Neither is this
assignment compatible with the monoidal structure on the category of
line operators. That is, it is not true, in general, that
$\frV_{\alpha\beta}(\sA,\sB)$ is isomorphic to
$$
\frV_{\alpha\beta}(\sA)\otimes \frV_{\alpha\beta}(\sB).
$$
This is in contrast with the situation in the GL-twisted theory
\cite{KW}.

The ultimate reason for this difference is that the twisted gauge
theory we are dealing with is not topological, but only
holomorphic-topological. Suppose we fix the location of the line
operator $\sB$, but vary the location of $\sA$ on $I\times C$. The
BRST-complex $\frV_{\alpha\beta}(\sA,\sB)$ is a differential graded
vector bundle over $I\times C$ with a connection along $I$ and a
$\bpartial$ operator along $C$. If one fixes $p\in C$ and varies
$y\in I$ (without colliding with $\sB$), then the BRST complexes are
all naturally isomorphic. But there is no isomorphism between
complexes corresponding to different $p$.

In the GL-twisted theory, one can choose all line operators to be at
the same point on $X_1$ and different points on $C$. Because line
operators are local along $C$, the supersymmetric quantum mechanics
describing this situation decomposes as a product of supersymmetric
quantum-mechanical systems corresponding to each line operator. This
implies that the quantum Hilbert space also factorizes.

In the holomorphic-topological field theory, if we want to study the
OPE, we have to work with all line operators inserted at the same
point on $C$ (but different points on $X_1$), and the arguments like
in the previous paragraph do not apply.

For simplicity, let us begin with the case where all line operators
are either Wilson or 't Hooft, with no ``mixed'' ones. When
quantizing the theory at weak coupling, the roles of Wilson and 't
Hooft operators are very different. 't Hooft operators directly
affect the equations for the BRST-invariant configurations whose
solutions determine the space of bosonic zero modes. A Wilson
operator corresponds to inserting an extra degree of freedom, which
couples weakly to the gauge fields, and can be treated
perturbatively.

The first step is to ignore the Wilson operators completely. As
explained in \cite{KW}, 't Hooft operators are line operators of
type $(B,A,A)$, i.e. they can be viewed either as line operators in
the B-model on $\cM_{Higgs}(G,C)$, or in the A-model on
$\cM_{flat}(G,\CC)$. When $\Sigma$ is flat and has no boundary, we
can regard the twisted gauge theory on $\Sigma\times C$ as a
supersymmetric sigma-model with $(4,4)$ supersymmetry. The
introduction of boundaries (either of A or B types) breaks $3/4$ of
supercharges and effectively eliminates one of the spatial
directions, so we end up with a supersymmetric quantum mechanics
with $N=2$ supersymmetry. The corresponding supersymmetry algebra
has a single complex supercharge $Q$ satisfying
$$
Q^2=0,\qquad \{Q,Q^\dag\}=2H,
$$
where $H$ is the Hamiltonian. $Q$-cohomology can be identified with
the space of supersymmetric ground states, i.e. states satisfying
$$
Qa=Q^\dag a=0.
$$
Strictly speaking, this is guaranteed only when the target space of
the supersymmetric quantum mechanics is compact. In the case of
interest to us, the target space is the Schubert cell $\cC_\mu$ (if
there is a single 't Hooft operator), or a product of several
Schubert cells, which are noncompact unless all coweights are
minuscule \cite{KW}. From the physical viewpoint, the correct
version of $Q$-cohomology is the $L^2$-cohomology, and we will
assume some version of Hodge theory works for the $L^2$-cohomology.

There are two well-known kinds of $N=2$ supersymmetric quantum
mechanics (SQM). $N=2$ SQMs of the first kind are classified by a
choice of a Riemannian target and a flat vector bundle $V$ over it;
its space of states is the space of differential forms with values
in $V$, and the corresponding operator $Q$ is the twisted de Rham
differential. This is the kind of effective SQM which appears when
considering 't Hooft operators as line operators in the A-model
\cite{KW}. It is clear that this SQM is not suitable for the
B-model, because once we include the Wilson operators, the bundle
over $\cC_\mu$ will not be flat. Also, in the B-model the BRST
operator $Q$ is likely to be a Dolbeault-type operator.

$N=2$ SQMs of the second kind look more promising: they are
classified by a choice of a K\"ahler target space and a holomorphic
vector bundle over it. The space of states is the space of
differential forms of type $(0,p)$ with values in a holomorphic
vector bundle $W$, and $Q$ acts as the Dolbeault operator.

In the next section we will perform the reduction to a SQM in some
detail and show that in the absence of Wilson operators $W$ is the
bundle of forms of type $(p,0)$ (for any $p$). But we can deduce
this result in a simpler way by making use of both A and B-models.
Indeed, if we take as our boundary conditions branes of type
$(B,A,A)$, we can interpret the space of ground states of the SQM in
terms of either model. For the Dolbeault cohomology of $W$ to be
isomorphic to the de Rham cohomology of $\cC_\mu$, $W$ has to be the
bundle
$$
\oplus_p \Omega^{p,0}(\cC_\mu).
$$
In the presence of a Wilson line, this also has to be tensored with
the holomorphic vector bundle corresponding to the Wilson line.

\subsection{Bosonic zero modes}
Our next task is to analyze bosonic zero modes in the presence of 't
Hooft operators. The BPS equations are simply the Bogomolny
equations, if the boundary conditions are suitably chosen \cite{KW}.
In fact, it has been shown in \cite{KW} that if in the absence of an
't Hooft operator the solution is unique, then in the presence of 't
Hooft operators the moduli space of solutions is $\cC_\mu$ (for a
single 't Hooft operator), or a tower of several Schubert cells
$\cC_{\mu_i}$ fibered over each other (for several 't Hooft
operators). So the bosonic zero modes span the tangent space to
$\cC_\mu$ or its generalization. However, it is useful to have an
explicit description of the tangent space in terms of solutions of
linearized Bogomolny equations in order to identify the fermionic
zero modes.

Recall that $w=y+ix_0$ with $y \in [0,1],\, x_0 \in \RR,$ is a
complex coordinate on $\Sigma$, while $z$ is a complex coordinate on
a closed Riemann surface $C$. For $t=i$ the BRST-invariant
``holomorphic connection'' on $\Sigma$ is \be {\cal
A}=(A_w+i\Phi_w)dw+(A_{\barw}+i\Phi_{\barw})d\barw \label{holcon}
\ee We further define the ``anti-holomorphic connection'' \be {\check
A}=(A_w-i\Phi_w)dw+(A_{\barw}-i\Phi_{\barw})d\barw \label{anticon}
\ee and introduce corresponding covariant differentials in the
adjoint representation:
$${\cal D}=\p +i[{\cal A},\cdot ]=
D-[\Phi,\cdot ],\quad {\check {\cal D}}=\p +i[{\check A},\cdot ]=D+[\Phi,\cdot ].$$
Note that holomorphic and anti-holomorphic connections are related
by Hermitean conjugation:
$$ {\cal A}^{\dg}={\check A}.$$

We set background $q_z$ and $\tq$ to zero. Then it can be shown analogously
to \cite{KW} that variations of these fields are also zero.
Therefore, the complete set of BPS equations is obtained by setting to zero the
BRST variations of gauginos. These are written down\footnote{In comparing with
\cite{htft} exchange $z$ and $w.$} in \cite{htft}.

Let us first consider one of the ``real'' BPS equations: \be -i\Bigl(D_w
\Phi_{\barw} +D_{\barw}\Phi_w\Bigr)=g_{w\barw}g^{z\barz} \Bigl(F_{z \barz}-
i[q_z,q_{\barz}]+2g_{z\barz}[\tq,\tq^{\dg}]\Bigr)
\label{odin} \ee where $w=y+ix^0$ and $z=x^1+ix^2.$ Variation of
(\ref{odin}) gives
 \be -i D_w \left(\delta \Phi_{\barw}\right)-i
D_{\barw} \left(\delta \Phi_{w}\right)+[\delta A_w,
\Phi_{\barw}]+[\delta A_{\barw}, \Phi_w]
=g_{w\barw}g^{z\barz}\Bigl(D_z \delta A_{\barz}-D_{\barz} \delta
A_z\Bigr) \label{vari} \ee where $2D_z=D_1-iD_2.$ We further assume
that all fields are independent of time $x^0$ and that background
fields $A_0=\Phi_0=0,$ so that $D_w=D_{\barw}=\half D_y$ and
$\Phi_w=\Phi_{\barw}=\half \Phi_y.$ Then, (\ref{vari}) becomes \be
-D_y(\delta \Phi_w +\delta\Phi_{\barw})+ i[\Phi_y,\delta A_w +
\delta A_{\barw}]+2ig_{w\barw}g^{z\barz} D_z(\delta
A_{\barz})-2ig_{w\barw}g^{z\barz} D_{\barz}(\delta A_z)=0
\label{varii} \ee

Now we impose a gauge-fixing condition: \be D_y\left(\delta {\cal
A}_w+\delta {\cal A}_{\barw}\right)+ [\Phi_y,\delta {\cal A}_w
+\delta {\cal A}_{\barw}]+ 4g_{w\barw}g^{z\barz} D_{z}(\delta
A_{\bar z})=0 \label{gauge} \ee

From (\ref{varii}) and (\ref{gauge}) follows \be
\frac{g^{w\barw}}{2}{\cal D}_y \left( \delta {\check A}_w +\delta
{\check A}_{\barw} \right) +2g^{z\barz} D_{\barz}\bigl(\delta
A_z\bigr)=0 \label{variii} \ee Taking hermitean conjugate of
(\ref{variii}) gives \be \frac{g^{w\barw}}{2}{\check {\cal D}}_y \left(
\delta {\cal A}_w +\delta {\cal A}_{\barw} \right) +2g^{z\barz}
D_z\bigl(\delta A_{\barz}\bigr)=0 \label{newvariii} \ee Next we
consider the complex BPS equations: \be {\cal F}_{\barz w}=0,\quad
{\cal F}_{\barz \barw}=0 \label{dva} \ee Variation of these two
equations gives \be D_{\barz} \left( \delta {\cal A}_w
\right)-{\cal D}_w(\delta A_{\barz})=0
\label{variv} \ee and \be D_{\barz} \left( \delta {\cal A}_{\barw}
\right)-{\cal D}_{\barw}(\delta A_{\barz})=0 \label{varv} \ee

The sum of (\ref{variv}) and (\ref{varv}) gives \be
D_{\barz}\left(\delta {\cal A}_w + \delta {\cal A}_{\barw}
\right)-{\cal D}_y\delta A_{\barz}=0
\label{varvi} \ee Taking hermitean conjugate of (\ref{varvi}) we
obtain \be D_z\left( \delta {\check A}_w +\delta {\check A}_{\barw}
\right)- {\check {\cal D}}_y\left(\delta A_z\right)=0 \label{varvii} \ee

We conclude that  $T{\cal M}$ splits into two parts. Holomorphic
bosonic modes from the first part satisfy Dirac-like equation: \be
O_1:=\left(
\begin{tabular}{cc}
${\cal D}_y$ & $ 2D_{\barz} $\\
$2 g^{z \barz} D_z$ & $-g^{w\barw}{\check {\cal D}}_y$\\
\end{tabular}
\right) \left(
\begin{tabular}{c}
$-\delta A_{\barz}$ \\
 $
\half\left(\delta {\cal A}_w +\delta {\cal A}_{\barw}\right)$ \\
\end{tabular}
\right)=0 \label{holrecast} \ee We impose boundary conditions \be
\delta A_{\barz}(0)=0,\quad \delta {\cal A}_w(1) +\delta {\cal
A}_{\barw}(1)=0 \label{boscond} \ee
The difference of (\ref{variv}) and (\ref{varv}) as well as
variation of the second ``real'' BPS condition ${\cal F}_{w \barw}=0$
give equations for the remaining holomorphic
bosonic variation $\delta {\cal A}_w -\delta{\cal A}_{\barw}$:
\be D_{\barz} \left(\delta {\cal A}_w -\delta {\cal
A}_{\barw}\right)=0, \quad
 {\cal D}_y\left(\delta {\cal A}_w -\delta {\cal A}_{\barw}\right)=0.
\label{otherhol} \ee Analogously,  ${\overline T}{\cal M}$ splits
into two parts. Some of the anti-holomorphic bosonic zero modes
 satisfy Dirac-like equation:
\be O_2:=\left(
\begin{tabular}{cc}
${\check {\cal D}}_y$ & $ 2D_z $\\
$2 g^{z \barz} D_{\barz}$ & $- g^{w\barw}{\cal D}_y$\\
\end{tabular}
\right) \left(
\begin{tabular}{c}
$-\delta A_z$ \\
 $\half\left(
\delta {\check A}_w +\delta {\check A}_{\barw}\right)$ \\
\end{tabular}
\right)=0 \label{antirecast} \ee We impose boundary conditions \be
\delta A_{z}(0)=0,\quad \delta {\check A}_w(1) +\delta {\check
A}_{\barw}(1)=0 \label{boscondii} \ee

The remaining anti-holomorphic bosonic variation $\delta {\check
A}_w -\delta{\check A}_{\barw}$ satisfy \be D_{z} \left(\delta
{\check A}_w -\delta {\check A}_{\barw}\right)=0, \quad
 {\check {\cal D}}_y\left(\delta {\check A}_w -\delta {\check A}_{\barw}\right)=0.
\label{otheranti} \ee

There are no non-trivial solutions of (\ref{otherhol}) and
(\ref{otheranti}) and we conclude that $T{\cal M}$(resp. ${\overline
T} {\cal M}$) is defined as the kernel of the operator $O_1$(resp.
$O_2$).

\subsection{Fermionic zero modes}
The gaugino equations of motion are: \be
D_z(\blam_{\barw})+D_{\barw}(\blam_z)+[\Phi_{\barw}, \lambda_z]=0
\label{fermi} \ee

\be D_z(\lambda_{w})+D_{w}(\lambda_z)+[\Phi_{w},\blam_z]=0
\label{fermii} \ee

\be D_w\blam_{\barw} +[\lambda_w,\Phi_{\barw}] -g_{w \barw} g^{z
\barz} D_{\barz } {\blam}_z=0 \label{fermiii} \ee

\be D_{\barw}\lambda_{w} +[\blam_{\barw},\Phi_{w}] -g_{w \barw} g^{z
\barz} D_{\barz } {\lambda}_z=0 \label{fermiv} \ee

The sum of (\ref{fermi}) and (\ref{fermii}) gives( recall
$\Phi_w=\Phi_{\barw}=\half \Phi_y,\quad D_w=D_{\barw}=\half D_y$)
\be 2 D_z(\lambda_w +\blam_{\barw})+ D_y(\lambda_z
+\blam_{\barz})+[\Phi_y,(\lambda_z +\blam_{\barz})]=0 \label{fermv}
\ee Meanwhile, the sum of (\ref{fermiii}) and (\ref{fermiv}) gives
\be 2 g^{z \barz} D_{\barz}(\lambda_z +\blam_{\barz}) -g^{w \barw}
D_y(\lambda_w +\blam_{\barw})+g^{w\barw} [\Phi_y,(\lambda_w
+\blam_{\barw})]=0 \label{fermvi} \ee

The two equations (\ref{fermv}) and (\ref{fermvi}) can be recast as
a Dirac-like equation: \be \left(
\begin{tabular}{cc}
${\check {\cal D}}_y$ & $ 2D_z $\\
$2 g^{z \barz} D_{\barz}$ &
$- g^{w \barw}{\cal D}_y$\\
\end{tabular}
\right) \left(
\begin{tabular}{c}
$\lambda_z +\blam_z$ \\
 $\lambda_w+\blam_{\barw}$ \\
\end{tabular}
\right)=0 \label{recastii} \ee Similarly, the difference of
equations (\ref{fermi}) and (\ref{fermii}) combines with the
difference of equations (\ref{fermiii}) and (\ref{fermiv}) into
another Dirac-like equation: \be \left(
\begin{tabular}{cc}
${\cal D}_y$ & $ 2D_z $\\
$2 g^{z \barz} D_{\barz}$ &
$- g^{w \barw}{\check {\cal D}}_y$\\
\end{tabular}
\right) \left(
\begin{tabular}{c}
$\blam_z-\lam_z$ \\
 $\blam_{\barw}-\lam_w$ \\
\end{tabular}
\right)=0 \label{morerecast} \ee We impose boundary conditions at
$y=0$ or $y=1$: \be \lambda_z(0)+\blam_z(0)=0,\quad
\lambda_w(1)+\blam_{\barw}(1)=0 \label{bry} \ee \be
\lambda_z(1)-\blam_z(1)=0,\quad \lambda_w(0)-\blam_{\barw}(0)=0
\label{morebry} \ee Note that (\ref{bry}) are BRST invariant
boundary condition, moreover they are BRST variations of the bosonic
boundary conditions (\ref{boscondii}). Meanwhile the BRST variation
of (\ref{morebry}) gives
$$q_z T \tq^{\dg}\vert_{y=1}=0$$
which is zero in the background we consider, i.e. with $q_z=0$ and
$\tq=0.$ Comparing (\ref{recastii}) with equations of motion for the
anti-holomorphic bosonic zero modes (\ref{antirecast}), we conclude
that  solutions of (\ref{recastii}) are in one-to-one correspondence
with elements of ${\overline T}{\cal M}.$

Eq. (\ref{morerecast}) has no nontrivial solutions for the following
reason. Let us denote by $O$ the operator in (\ref{morerecast}). In
addition to (\ref{morebry}) we impose boundary conditions on ghost
number $-1$ fermions \be \lam_{\barz w}(0)-\blam_{\barz
\barw}(0)=0,\quad \lam_{w \barw}(1)-\blam_{w \barw}(1)=0
\label{morebryii} \ee The boundary conditions (\ref{morebry}) and
(\ref{morebryii}) are chosen so that in computing hermitean
conjugate ${\cal O}^{\dg}$ we can drop boundary terms obtained from
integration by parts. Then  we find \be O^{\dg}
O=-\Bigl(2\Delta_C+\half \Delta_{\Sigma}\Bigr)I_{2\times 2}
\label{proof} \ee where $\Delta_C=g^{z
\barz}(D_{\barz}D_z+D_zD_{\barz})$ and $\Delta_{\Sigma}=g^{w
\barw}({\check {\cal D}}_y{\cal D}_y+{\cal D}_y {\check {\cal D}}_y).$ In
obtaining (\ref{proof}) we used BPS equations for the background
fields. Since both $-\Delta_C$ and $-\Delta_{\Sigma}$ are
nonnegative operators, the kernel of the operator $O$ must be
annihilated by both Laplacians. This implies, in particular, that
$\blam_z-\lam_z$ is constant on the interval $y \in[0,1].$ However,
such a mode is necessarily zero due to boundary conditions
(\ref{morebryii}).

Equations of motion for matter fermions (using the $\cN=2$ language)
are \be \left(
\begin{tabular}{cc}
${\check {\cal D}}_y$ & $2  D_{\barz}$ \\
$2 g^{z \barz} D_{z}$ & $-g^{w \barw}{\cal D}_y$\\
\end{tabular}
\right) \left(
\begin{tabular}{c}
$\psi_{\barz} +\bchi_{\barz}$ \\
$-\left(\psi_{\barw}+\bchi_w\right)$ \\
\end{tabular}
\right)=0 \label{recastiii} \ee and \be \left(
\begin{tabular}{cc}
${\cal D}_y$ & $2D_{\barz}$ \\
$2 g^{z \barz} D_{z}$ & $- g^{w \barw}{\check {\cal D}}_y$\\
\end{tabular}
\right) \left(
\begin{tabular}{c}
$\bchi_{\barz}-\psi_{\barz}$ \\
$\psi_{\barw}-\bchi_w$ \\
\end{tabular}
\right)=0 \label{recastiiii} \ee

We impose the following boundary conditions at $y=0$ or $y=1$: \be
\psi_{\barw}(1) -\bchi_{w}(1)=0,\quad
\bchi_{\barz}(0)-\psi_{\barz}(0)=0 \label{bryii} \ee \be
\psi_{\barw}(0) +\bchi_{w}(0)=0,\quad
\bchi_{\barz}(1)+\psi_{\barz}(1)=0 \label{newbryii} \ee Note that
(\ref{newbryii}) are BRST invariant, meanwhile the BRST variation of
(\ref{bryii}) gives
$$D_{\barz}\tq^{\dg}\vert_{y=0}=0,\quad
{\cal D}_y \tq^{\dg}\vert_{y=1}=0
$$

Matter fermions (\ref{recastiiii}) belong to $T{\cal M},$ as can be
seen by comparing with (\ref{holrecast}). Eq. (\ref{recastiii}) has
no nontrivial solutions. The proof is similar to that for operator
$O$ above. Let us denote by $O'$ the operator in (\ref{recastiii}).
In addition to (\ref{newbryii}) we impose boundary conditions on
ghost number $-1$ fermions \be \chi_{z \barw}(0)+\bpsi_{z
w}(0)=0,\quad \chi_{z \barz}(1)+\bpsi_{z \barz}(1)=0
\label{newmorebryii} \ee The boundary conditions (\ref{newbryii})
and (\ref{newmorebryii}) are chosen so that when computing hermitean
conjugate ${O'}^{\dg}$ we can drop boundary terms obtained from
integration by parts. Then  we use BPS equations for the background
to show \be {O'}^{\dg} O'=-\Bigl(2\Delta_C+\half
\Delta_{\Sigma}\Bigr)I_{2\times 2} \label{proofii} \ee where
$\Delta_C=g^{z \barz}(D_{\barz}D_z+D_zD_{\barz})$ and
$\Delta_{\Sigma}=g^{w \barw}({\check {\cal D}}_y{\cal D}_y+{\cal D}_y
{\check {\cal D}}_y).$ Since both $-\Delta_C$ and $-\Delta_{\Sigma}$ are
non-negative operators, the kernel of the operator $O'$ must be
annihilated by both Laplacians. This implies, in particular, that
$\bchi_{\barz}+\psi_{\barz}$ is constant on the interval $y
\in[0,1].$ However, such a mode is necessarily zero due to boundary
conditions (\ref{newbryii}).

The result of this analysis is that fermionic zero modes span
$T\cM\oplus {\overline T}\cM$. Therefore the Hilbert space of the
effective SQM is the space of $L^2$ sections of the vector bundle
$$
\oplus_p \Lambda^p\left(T^*\cM\oplus {\overline
T^*}\cM\right)=\oplus_{p,q}\Omega^{p,q}(\cM).
$$
From the formulas for BRST transformation we see that BRST variation
of bosonic zero modes are precisely the fermionic zero modes
spanning ${\overline T}\cM$, while BRST variations of fermionic zero
modes vanish. This means that the BRST operator acts as the
Dolbeault operator.

\section{OPE of Wilson-'t Hooft operators for $G=PSU(2)$}

In this section we study in detail the OPE of WH loop operators in
the special case $G=PSU(2)$. The main goal is to test the
predictions of S-duality explained in \ref{s:sduality}.

\subsection{OPE of a Wilson and an 't Hooft operator}

Let us begin by considering the OPE of a Wilson and an 't Hooft
operator. The most naive approach is to regard an 't Hooft operator
as creating a classical field configuration, and analyze the
electric degree of freedom corresponding to the Wilson operator in
this classical background. As explained above, the field singularity
at the insertion point of an 't Hooft operator $T_\mu$ breaks the
gauge group $G=PSU(2)$ down its subgroup $H=U(1)$, so it seems that
all we have to do is to decompose the representation $R$ associated
to the Wilson operator into irreducibles with respect to $H$. If we
label representations of $PSU(2)$ by an even integer $n$ which is
twice the isospin, and denote the magnetic charge of the 't Hooft
operator  by $m\in \NN$, then the OPE at weak coupling appears to be
$$
T_m \cdot W_n = WT_{m,n}+WT_{m,n-2}+\ldots +WT_{m,-n}.
$$
But this contradicts S-duality, which requires that there be a
symmetry under $n\ra -m, m\ra n$. In fact, S-duality predicts that
the OPE also contains contributions from WH operators with smaller
magnetic charge. As explained in \cite{KW}, this is due the
``monopole bubbling'': the magnetic charge of an 't Hooft operator
can decrease by $2$ when it absorbs a BPS monopole. Such process is
possible because the moduli space of solutions of the Bogomolny
equations is noncompact for $m>1$; configurations with smaller
magnetic charge can be associated with points at infinity. The naive
argument ignored monopole bubbling and therefore missed all such
contributions.

This explanation also suggests that for $m=1$, where the moduli
space is simply $\PP^1$ and therefore is compact, the naive argument
is valid. To compare this with the S-duality predictions, we follow
the procedure outlined in section \ref{s:sduality}. To the loop
operators $T_1$ and $W_n$ one associates Laurent polynomials
$$
WT_{1,0}(x)=x+x^{-1},\quad WT_{0,n}(y)=y^n+y^{n-2}+\ldots + y^{-n}.
$$
To the WH loop operator $WT_{1,k}$ one associates the Laurent
polynomial
$$
WT_{1,k}(x,y)=x y^k+x^{-1}y^{-k}.
$$
We see that
\begin{equation}\label{opeone}
WT_{1,0}(x)WT_{0,n}(y)=\sum_{j=0}^n WT_{1,n-2j}(x,y),
\end{equation}
in agreement with the naive formula.

This example also provides a nice illustration of the difference
between line and loop operators. Recall that the Hilbert space
$\frV(A)$ associated to the line operator $WT_{1,k}$ is the space of
sections of the differential graded vector bundle
$$
\cO(-k)\otimes \oplus_{p,q} \Omega^{p,q}
$$
over the Schubert cell $\cC_\mu\simeq \PP^1$, with the differential
being the Dolbeault differential. Here the first factor comes from
the electric degree of freedom, and the rest comes from fermionic
zero modes. Instead of this differential graded vector bundle, we
can think of the corresponding coherent sheaf
$$
\cO(-k)\otimes \Omega^*(\PP^1)=\cO(-k)+\cO(-k-2).
$$
The SQM Hilbert space is the Dolbeault resolution of this coherent
sheaf, so instead of thinking about the BRST cohomology, we can
think about the cohomology of this sheaf. Thus the sum of the WH
line operators on the right-hand side of eq. (\ref{opeone})
corresponds to the coherent sheaf
$$
\left(\cO(-n)+\cO(-n+2)+\ldots +\cO(n)\right)\otimes
\Omega^*(\PP^1).
$$
On the other hand, the product of a Wilson operator $W_n$ and an 't
Hooft operator $T_1$ gives a trivial vector bundle of rank $n+1$
over $\PP^1$, tensored with $\Omega^*(\PP^1)$. Clearly, the equality
between left-hand side and right-hand side of eq. (\ref{opeone})
does not hold on the level of line operators, because
\begin{equation}\label{ineq}
\cO\otimes \CC^{n+1}\neq \cO(-n)+\cO(-n+2)+\ldots +\cO(n).
\end{equation}
But the equality does hold on the level of K-theory.\footnote{We are
grateful to Roman Bezrukavnikov for providing the following
argument.} To see this, we will exhibit a filtration of $\cO\otimes
\CC^{n+1}$ whose cohomology is precisely the right-hand-side of eq.
(\ref{ineq}). Recall that $\PP^1=G_\CC/B$, where $G_\CC=SL(2,\CC)$
and $B$ is the group of upper-triangular matrices with unit
determinant. The fiber of the trivial vector bundle $V$ of rank
$n+1$ carries the representation of $G_\CC$ of isospin $n$; for
example, we can realize it by thinking of the fiber of $V$ as the
space of homogeneous degree-$n$ polynomials in variables $u$ and
$v$, which we denote $\cD_n(u,v)$. $SL(2,\CC)$ acts on it by linear
substitutions. To define a filtration on $V$, we can specify a
$B$-invariant filtration on $\cD_n(u,v)$. The obvious filtration is
to take $F_k$ to be the subspace of $\cD_n(u,v)$ consisting of
polynomials of degree $k$ or lower in $u$, with $k$ ranging from $0$
to $n$. It is easy to check that $F_k$ is $B$-invariant for any $k$.
Obviously, $F_k/F_{k-1}$ is one-dimensional and the maximal torus of
$SL(2,\CC)$ acts on it with weight $2k-n$. Hence $V$ acquires a
filtration of length $n+1$ whose $k$-th cohomology is $\cO(2k-n)$.

\subsection{OPE of WH operators with minuscule coweights}

In this subsection we compute the product of WH with the smallest
nontrivial coweights (for $G=PSU(2)$). The weights may be arbitrary.
This case is very special, because when the WH operators are not
coincident, the moduli space of Bogomolny equations is compact. This
happens because the smallest nontrivial coweight of $G=PSU(2)$ is
minuscule.\footnote{The corresponding representation of $\LG=SU(2)$
has the property that all its weights lie in a single Weyl orbit.}
Therefore the monopole bubbling is absent, as discussed in
\cite{KW}. The main difficulty is to determine the behavior of the
zeromode wavefunctions in the limit when the two WH operators
coincide.

Let us recall what the moduli space of Bogomolny equations looks
like for two noncoincident WH operators with $\mu=1$ located at the
same point on $C$ \cite{KW}. It is a Hirzebruch surface $F_2$ which
is a fibration of $\PP^1$ over $\PP^1$. One can think of it as a
blow-up of the weighted projective plane $\WP_{112}$ at the
$\ZZ_2$-orbifold point. This blow-up is associated with moving the
WH operators apart in the $y$ directions. Thus $\WP_{112}$ is the
coincidence limit of the moduli space. The orbifold point
corresponds to the trivial solution of the Bogomolny equations
(without the monopole singularity), while the complement of the
orbifold point is isomorphic to $T\PP^1$ and corresponds to
solutions of the Bogomolny equations with one singularity of
coweight $\mu=\pm 2$. This implies \cite{KW} that the product of two
WH operators with coweight $\mu=\pm 1$ may contain WH operators with
coweight $\mu=\pm 2$ and WH operators with coweight $\mu=0$. To
understand which WH operators appear in the product, one has to
understand the zeromode wavefunctions in the coincidence limit.
Those which remain spread-out on the complement of the orbifold
point correspond to WH operators with $\mu=\pm 2$, while those which
concentrate in the neighborhood of the exceptional divisor
correspond to WH operators with $\mu=0$.

As explained above, the wavefunctions of the effective SQM in the
presence of WH operators are square-integrable forms on the moduli
space with values in a certain holomorphic line bundle which satisfy
the equations
\begin{equation}\label{eqsdolbope}
{\bar D}\rho=0,\qquad {\bar D}^\dag\rho=0.
\end{equation}
where ${\bar D}$ is the covariant Dolbeault differential\footnote{
See sections 6.3 and 6.4 for appropriate ${\bar D}.$}.
In the coincidence limit, the K\"ahler metric on the moduli space
degenerates so that in the neighborhood of the orbifold point it
becomes a flat metric on $\CC^2/\ZZ_2$. More generally, when the WH
operators are close to each other, the metric in the neighborhood of
the exceptional divisor is well-approximated by a hyperk\"ahler
metric on the blow-up of $\CC^2/\ZZ_2$ \cite{KW}. This is because
this region in the moduli space corresponds to solutions of the
Bogomolny equations which are trivial everywhere except in a small
neighborhood of a point on $C\times I$ (the point at which one of
the WH operators is inserted). Such solutions are arbitrarily well
approximated by patching together solutions on $\RR^3$ with the
trivial solution on $C\times I$. Therefore the metric will be
arbitrarily well approximated by the metric on the moduli space of
Bogomolny equations on $\RR^3$, which is hyperk\"ahler.

The blow-up of $\CC^2/\ZZ_2$ is isomorphic to $T^*\PP^1$ and has a
unique asymptotically flat hyperk\"ahler metric: the Eguchi-Hanson
metric. Therefore, one can produce approximate solutions of
equations (\ref{eqsdolbope}) on $F_2$ by first solving them on the
Eguchi-Hanson space and on $T\PP^1$, assuming square-integrability
in both cases, and patching them with the zero solution on the
remainder of $F_2$. The solutions coming from the Eguchi-Hanson
space will represent contributions to the zeromode Hilbert space
from WH operators with $\mu=0$, while the solutions coming from
$T\PP^1$ will represent contributions from $\mu=\pm 2$.

The contribution to the product of WH operators coming from $T\PP^1$
will be called the ``bulk'' contribution, while the one coming from
$T^*\PP^1$ will be called the ``bubbled'' contribution, because it
is due to monopole bubbling. The ``bulk'' contribution is rather
trivial and in fact can be determined without any computations: the
magnetic charges of the singularities simply add up, the same
applies to the electric charges, and therefore the bulk contribution
must be simply
$$
WT_{2,2m+2k}.
$$
The ``bubbled'' contributions are much more subtle and will be
determined below by solving the equations (\ref{eqsdolbope}) on
$T^*\PP^1$. We will also solve the same equations on $T\PP^1$, not
because it is required to determine the operator product, but
because this computation will provide a consistency check on our
approach, see section 6.5.

As a preliminary step, let us exhibit the predictions of $SL(2,\ZZ)$
duality for the product of WH operators with coweight $\mu=\pm 1$:
$$
WT_{1,2m}\cdot
WT_{1,2k}=WT_{2,2m+2k}+WT_{0,2m-2k}-WT_{0,0}-WT_{0,2m-2k-2}
$$
Here $m$ and $k$ are integers, and we assume $m\neq k$. We can
simplify our problem a bit by noting that by applying the
$T$-transformation several times, we can reduce to the case $k=0$,
in which case the duality predicts that for $m\neq 0$ we have
\begin{equation}\label{maineqtotest}
WT_{1,2m}\cdot WT_{1,0}=WT_{2,2m}+WT_{0,2m}-WT_{0,0}-WT_{0,2m-2}
\end{equation}
The ``bulk'' contribution is as expected, while the ``bubbled''
contributions are far from obvious. Note that some of the
coefficients are negative, unlike for 't Hooft operators in
\cite{KW}. This is because we are working in the K-theory of the
category of line operators, where negative signs occur naturally.

Similar manipulations in the case $m=0$ lead to
\begin{equation}\label{opeKW}
T_1\cdot T_1=T_2+T_0.
\end{equation}
This is S-dual to the fact that the tensor square of the defining
representation of $SU(2)$ is a sum of the adjoint representation
(corresponding to the 't Hooft operator $T_2$) and the trivial
representation (corresponding to $T_0$). This prediction was checked
in \cite{KW} for the GL-twisted theory. Briefly speaking, in the
GL-twisted theory we are looking for harmonic square-integrable
forms on $F_2$ and study their behavior in the limit when $F_2$
degenerates to $\WP_{112}$. Since topologically $F_2$ is the same as
$\PP^1\times\PP^1$, and harmonic forms can be interpreted in
topological terms (as cohomology classes), we know a priori that the
dimension of the space of harmonic forms is the same as the
dimension of $H^*(\PP^1\times\PP^1,\CC)$, which is four. It is also
well-known that there is a unique square-integrable harmonic form on
the Eguchi-Hanson space (in degree 2). Therefore the Eguchi-Hanson
space contributes one state, and $T\PP^1$ contributes three states.
The latter states arise precisely from the quantization of the
moduli space of the Bogomolny equations with a single singularity of
coweight $\mu=\pm 2$. This leads to the formula (\ref{opeKW}), as
predicted by S-duality.

The case $m\neq 0$ is different in two respects. First of all, we
have to consider forms with values in a holomorphic line bundle
$\cL$. Second, the equations we have to solve (\ref{eqsdolbope})
involve the Dolbeault operator rather than the de Rham operator.

To fix $\cL$, let us use the same boundary conditions as before,
i.e. assume that the boundary condition on which $WT_{1,2m}$ acts
corresponds to a particular $PSU(2)$ bundle on $C$. Then the line
bundle on $F_2$ is the pull-back of $\cO(-2m)$ from the base
$\PP^1$. (This is because the electric degree of freedom is
associated, via weight $2m$, with the $U(1)$ bundle coming from the
first Hecke transformation and does not care about the second Hecke
transformation. The base $\PP^1$ is the parameter space for the
first Hecke transformation, while the fiber $\PP^1$ is the parameter
space for the second Hecke transformation.) Therefore, in the
``bulk'' part of the computation, $\cL$ is simply the pull-back of
$\cO(-2m)$ from the base of $T\PP^1$ to the total space.

Similarly, in the ``bubbled'' part of the computation the line
bundle is a pull-back of $\cO(-2m)$ from the base of $T^*\PP^1$ to
the total space. To see this, we can make use of an explicit
description of $F_2$ as a K\"ahler quotient of $\CC^4$ by $U(1)^2$
\cite{KW}. Let the coordinates on $\CC^4$ be $u,v,b,$ and $b'$. The
first $U(1)$ action has weights $1,1,2,0$, and the second $U(1)$
action has weights $0,0,1,1$. The moment map equations are
$$
|u|^2+|v|^2+2|b|^2=1,\quad |b|^2+|b'|^2=d.
$$
where $d$ is assumed to be positive and smaller than $1/2$. These
equations imply that $u$ and $v$ cannot vanish simultaneously and
can be regarded as homogeneous coordinates on $\PP^1$. Therefore the
map $(u,v,b,b')\mapsto (u,v)$ defines a fibration over $\PP^1$. Its
fiber is also a $\PP^1$ with homogeneous coordinates $b$ and $b'$.
To degenerate $F_2$ into $\WP_{112}$ one need to take the limit
$d\ra 1/2$. The exceptional divisor is given by $b'=0$. The
neighborhood of the exceptional divisor is the subset given by
$b\neq 0$. We can see that it is a copy of $T^*\PP^1$ by letting
$a=b'/b$. Since $u,v,a$ have zero weights with respect to the second
$U(1)$ and since every orbit of the second $U(1)$ action contains a
unique representative with ${\rm arg}(b)=0$, we conclude that the
subset $b\neq 0$ can be identified with the K\"ahler quotient of
$\CC^3$ parameterized by $u,v,a$ by the first $U(1)$. Since the
weights of these variables are $1,1,-2$, and $u$ and $v$ cannot
vanish simultaneously, this quotient is the total space of the line
bundle $\cO(-2)$ over $\PP^1$, which is the same as $T^*\PP^1$.

Now, the line bundle $\cL$ on $F_2$ can be defined as the quotient
of the space of quintuples $u,v,b,b',\rho$ by the $(\CC^*)^2$ action
with weights $$(1,0),(1,0),(2,1),(0,1),(-2m,0).$$ The variable
$\rho$ parameterizes the fiber of $\cL$. When we restrict to the
subset $b\neq 0$, we may forget about the second $\CC^*$, and
replace $b$ and $b'$ with $a=b'/b$. Thus the restriction of $\cL$ to
this subset is the quotient of the space of quadruples $u,v,a,\rho$
by the $\CC^*$ action with weights $1,1,-2,-2m$. This is clearly the
total space of the line bundle over $T^*\PP^1$ which is a pull-back
of $\cO(-2m)$ on the $\PP^1$ base.

\if To count observables we decompose ${\cal M}$ into ``bulk part''
$TP^1,$ which captures the vicinity of the curve $C_1=s+f$ with
normal bundle\footnote{ The degree of the normal bundle to a curve
in the class $C$ is given by $-C\cdot K.$} $O(2)$ and ''bubbled
part'' $T^*P^1,$ which zooms at the vicinity of the curve $C_2=s-f$
with normal bundle $O(-2).$ Both $C_1$ and $C_2$ are $P^1$'s as
follows from adjunction formula for the genus of holomorphic curve
$C$ inside complex surface with canonical class $K=-2s-4f.$
$$2g-2=C^2+C\cdot K.$$
Therefore, to check (\ref{basicii}) we should compute
$H^p\left(\Omega^q\otimes O(2m+2k),TP^1\right)$  and
$H^p\left(\Omega^q\otimes O(2m-2k),T^*P^1\right).$  In the following
two subsections we compute these cohomologies. \fi

\subsection{Wavefunctions on $T\PP^1.$}

Let $u,v,b$ be homogeneous coordinates on $T\PP^1$, with $\CC^*$
weights $1,1,2$. Let us work in the patch $u\neq 0$ and define
inhomogeneous ``coordinates'' \be z=\frac{v}{u},\quad
w=\frac{\sqrt{b}}{u} \label{defi} \ee We put the word
``coordinates'' in quotation marks, because $w$ is defined up to a
sign and is not really a good coordinate. The good coordinate is
$w^2$.

Our goal is to solve equations (\ref{eqsdolbope}) on $T\PP^1$, i.e.
to find harmonic representatives of the $L^2$ Dolbeault cohomology
groups
$$
H^p(\Omega^q(-2m),T\PP^1),\qquad p,q,=0,1,2.
$$
Here
$$
\Omega^q(-2m)=\Omega^q\otimes \cO(-2m).
$$
The sum of these cohomology groups is nothing but the vector space
$\frV(\sA)$, where $\cA$ is the WH operator $WT_{2,2m}$. In section 6.5 we
will use the knowledge of $\frV(\sA)$ for this and other WH
operators on the r.h.s. of eq. (\ref{maineqtotest}) to make a
consistency check on our computations.

\subsubsection{The metrics}

\if $\chi\bigl(TP^1,O(2m)\bigr)=\sum_{p=0}^2\sum_{q=0}^2 (-)^{p+q}
h^p\bigl(\Omega^q(2m),TP^1\bigr)$ where $\Omega^q(k)=\Omega^q
\otimes O(k).$ \fi

While we do not know the precise form of the K\"ahler metric on
$T\PP^1$ coming from the Bogomolny equations, it is tightly
constrained by symmetry considerations. Indeed, $PSU(2)$ gauge
transformations act on the moduli space by isometries which preserve
the complex structure, and the orbits have real codimension $1$,
therefore the most general ansatz will depend on functions of a
single variable. The $PSU(2)$ action in question acts on $u,v$ as a
two-dimensional projective representation, and acts trivially on
$b$. Using this, it is easy to show that the most general
$PSU(2)$-invariant $(1,1)$-form on $TP^1$ is
$$J=f_1(\lambda) e_1 \wdg \bare_1+ f_2(\lambda) e_2 \wdg \bare_2$$
where \be e_1=\frac{dw}{w}-\barz e_2,\quad e_2= \frac{dz}{1+\vert z
\vert^2} \label{defii} \ee and $f_1,f_2$ are functions of the
$PSU(2)$ invariant \be \lambda=\frac{\vert w \vert^2}{1+\vert
z\vert^2} \label{defiii} \ee The K\"ahler condition $dJ=0$ implies
$f_1=-\lambda f_2'$ so that geometry is specified in terms of a
single function $f_2(\lambda)$ on $[0,\infty)$. Its behavior at zero
is constrained by the requirement that the metric be smooth at
$w=0$. Its behavior at infinity is constrained by the requirement
that after one-point compactification of $T\PP^1$ the neighborhood
of infinity looks like $\CC^2/\ZZ_2$ with a flat metric. These two
conditions are equivalent to

\be f_2\rightarrow \frac{1}{2\lambda} \quad for \quad
\lambda\rightarrow \infty,\quad f_2\rightarrow const \quad for \quad
\lambda \rightarrow 0. \label{asymp} \ee

The standard Fubini-Study metric on $\WP_{112}$ corresponds to
specific $f_1,f_2$ with these asymptotics:
$$f_1=\frac{(\sqrt{1+8\lambda^2}-1)^2}{4 \lambda^2 \sqrt{1+8\lambda^2}},
\quad f_2=\frac{\sqrt{1+8\lambda^2}-1}{4 \lambda^2}$$

Let us consider the line bundle $\cO(n)$ over $T\PP^1$. The $PSU(2)$
action on $T\PP^1$ lifts to a $PSU(2)$ action on $\cO(n)$ if $n$ is
even, or to an $SU(2)$ action if $n$ is odd. We are mainly
interested in even $n$. In a unitary trivialization, the most
general $SU(2)$-invariant connection on $\cO(n)$ is: \be
 A^{(n)}=\frac{\lambda f'_{(n)}}{f_{(n)}}e_1-\frac{n}{2}\barz e_2,\quad {\overline A}^{(n)}=-\frac{\lambda f'_{(n)}}{f_{(n)}}\bare_1+
\frac{n}{2}z \bare_2 \label{conn} \ee and covariant differentials
are defined as
$$ D=\p+A^{(n)},\quad \Delbar=\delbar +{\overline A}^{(n)}$$
For $n=-2m,\, m\in \ZZ$ the function  $f_{(-2m)}$ has the following
asymptotics:
\be f_{(-2m)} \rightarrow \lambda^{m} \quad
for \quad \lambda\rightarrow \infty,\quad f_{(-2m)}\rightarrow 1
\quad for \quad \lambda \rightarrow 0. \label{asympii} \ee The
asymptotic at $\lambda\rightarrow \infty$ is chosen in such a way
that the norm of the holomorphic section $w^{-2m}$ approaches a
constant, i.e. we go to the unitary trivialization \be
s_{unit}=s_{hol}(1+|z|^2)^mf_{(-2m)}(\lambda) \label{transf} \ee and
require the pointwise norm $\vert s_{unit} \vert^2$ to approach a
constant. The reason is that in the neighborhood of the orbifold
point $u=v=0$ $w^{-2m}$ represents a section which transforms
trivially between the two charts $u\ne 0$ and $v\ne 0,$ and provides
a local holomorphic trivialization of $\cO(-2m)$. We would like its
norm neither to diverge nor to become zero at the orbifold point.
The postulated behavior at $\lambda \rightarrow 0$ ensures that the
connection is smooth at the zero section of $T\PP^1$.

\subsubsection{A heuristic argument}

Since solving partial differential equations is hard, it is useful
to have some idea about the kind of solutions one expects to find.
There is a heuristic argument, explained to us by Roman
Bezrukavnikov, which gives the dimensions of the cohomology groups
we are after. Let us start with the case $m=0$ where we already know
the structure of solutions \cite{KW}: all of cohomology is of type
$(p,p)$, and there is a single solution for $p=0,1,2$:
$$
H^p(T\PP^1,\Omega^q)=\delta_{pq}V_1
$$
where $V_{2j+1}$ stands for $(2j+1)-$dimensional
irreducible representation of $SL(2,\CC).$

Next we note that the line bundle $\cO(2)$ corresponds to the
divisor $D$, where $D$ is the zero section of $T\PP^1$. Hence we
have a short exact sequence of coherent sheaves on $T\PP^1$:
$$
0\ra \cO\ra \cO(D)\ra N_D\ra 0,
$$
where $N_D$ is the normal bundle of $D$. This gives a long exact
sequence for sheaf cohomology groups. We are of course interested
not in sheaf cohomology groups, but in $L^2$ Dolbeault cohomology of
the corresponding line bundles. But let us cheat and ignore this
distinction. Then the long exact sequence implies
$$
H^0(T\PP^1,\cO(2))=V_1+V_3, \quad H^i(T\PP^1,\cO(2))=0,\ i=1,2.
$$

Similarly, if we tensor the short exact sequence with the sheaf
$\Omega^i$, $i=1,2$, and then write down the corresponding long
exact sequences, we infer:
$$
H^0(T\PP^1,\Omega^1(2))=V_1, \quad H^0(T\PP^1,\Omega^2(2))=0, \quad
H^i(T\PP^1,\Omega^j(2))=0,\ i,j=1,2.
$$

Having determined all relevant cohomology groups for $m=-1$, we can
move on to $m=-2$ and write down a short exact sequence involving
$\cO(4)\simeq \cO(2D)$:
$$
0\ra \cO(D)\ra \cO(2D)\ra N_D(D),
$$
which implies
\begin{equation}
H^0(T\PP^1,\cO(4))=V_1+V_3+V_5,\quad H^0(T\PP^1,\Omega^1(4))=V_3+V_1+V_3,
\end{equation}
$$H^0(T\PP^1,\Omega^2(4))=V_1,$$
with all higher cohomologies vanishing. Continuing in this fashion,
we can determine cohomology groups for all negative $m$. We find
that only degree-0 cohomology is nonvanishing. If we let $k=-m>0$,
then degree-0 cohomology groups are
\begin{eqnarray}
&H^0(T\PP^1,\cO(2k))=\sum_{j=0}^k V_{2j+1},\\
&H^0(T\PP^1,\Omega^1(2k))=V_{2k-1}+\sum_{j=1}^{k-1}V_{2j-1}+
\sum_{j=1}^{k-1}V_{2j+1},\\
&H^0(T\PP^1,\Omega^2(2k))=\sum_{j=0}^{k-2}V_{2j+1}.
\end{eqnarray}

If $m>0$, we can find  cohomology groups by applying
Kodaira-Serre duality to the results for $m<0$:
$$
H^p(T\PP^1,\Omega^q(2m))=H^{2-p}(T\PP^1,\Omega^{2-q}(-2m)).
$$
Thus for positive $m$ only degree-$2$ cohomology is nontrivial.

Below we write down an explicit basis for degree-0 cohomology groups
and check that all elements of the basis are square-integrable. By
Kodaira-Serre duality, this also verifies the predictions for
degree-$2$ cohomology. We have not been able to prove that
degree-$1$ $L^2$ cohomology groups vanish for all $m$. We only
checked that degree-$1$ $L^2$ cohomology, if it exists, does not
contain irreducible $PSL(2,\CC)$ representations of dimensions $1$
and $3$. (For larger $PSL(2,\CC)$ representations, the analysis
becomes very complicated, and we were not able to push it through.)

\subsubsection{$H^0(T\PP^1,\cO(2k)).\quad k>0.$} First we find holomorphic sections
of the line bundle $\cO(2k)$ on $T\PP^1$ . In a holomorphic
trivialization these sections are $ b^{k-j}P_{2j}(u,v)$ for
$j=0,\ldots, k$ where $P_{2j}(u,v)$ is a homogeneous polynomial of
degree $2j$ in variables $u,v$. For each $j$ these sections
transform in a representation $V_{2j+1}.$

To see that all these sections are
in $L^2$ we go to the unitary trivialization (\ref{transf}) and
compute the norm. For $s_{hol}=w^{2(k-j)} z^p$ with $p\le 2j \le 2k$
we find \be \int_{T\PP^1}\vert s_{unit} \vert^2 \,
f_1(\lam)f_2(\lam) e_1\wdg \bare_1 \wdg e_2 \wdg \bare_2=
\frac{\pi}{2}\int \frac{d\lam}{\lam} \lam^{2(k-j)}\, f_1\, f_2\,
f_{(2k)}^2
 \, \int \frac{\vert z \vert^{2p} dz d\barz}{\bigl(1+\vert z\vert^2\bigr)^{2+2j}}
\label{norma} \ee where we used (\ref{defii}) and (\ref{transf}).
The $z$-integral is convergent for $p\le 2j$, while the integral
over $\lam$ is finite since the integrand behaves \footnote{We use
the asymptotics (\ref{asymp}) and (\ref{asympii}) of
$f_1,f_2,f_{(2k)}$.} at infinity as $\frac{1}{\lam^{2j+3}}$ and at
zero as $\lam^{1+2(k-j)}$ for $j=0,\ldots,k.$

\subsubsection{$H^0(T\PP^1,\Omega^1(2k)),\quad k>0.$} Next we find holomorphic
sections of the vector bundle $\Omega^1(2k)$ on $T\PP^1$. Again it
is easy to do it in a holomorphic trivialization. $2k-1$ sections
pulled back from the base transform in a representation $V_{2k-1}$:
$$\rho_{hol}=uvP_{2k-2}(u,v)\bigl({du\over u}-{dv\over v}\bigr).$$
All these sections are square-integrable. Indeed, in the chart
$u\neq 0$ they are of the form $\rho_{hol}=z^pdz$ for
$p=0,\ldots,2k-2$ and their norm is finite: \be \rho_{unit}\wdg
{\overline
*\rho_{unit}}= {\pi \over 2}\int {d\lam \over \lam} f_1\,
f_{(2k)}^2
 \, \int {\vert z \vert^{2p} dz d\barz \over \bigl(1+\vert
 z\vert^2\bigr)^{2k}}<\infty .
\label{normaii} \ee

Also, there are holomorphic sections of the form
$$\rho_{hol}=\bigl(db-b{du\over u}-b{dv \over v}\bigr)F_{2k-2}(u,v,b)+
b\F_{2k-2}(u,v,b)\bigl({du\over u}-{dv\over v}\bigr),$$ where
$F_{2k-2}$ and $\F_{2k-2}(u,v,b)$ must satisfy(to ensure
non-singular behavior)
$$\F_{2k-2}-F_{2k-2}=ug_{2k-3}(u,v,b),\quad
\F_{2k-2}+F_{2k-2}=v\tg_{2k-3}(u,v,b).$$
We further  write
$$g_{2k-3}(u,v,b)=\sum_{j=1}^{k-1} b^{k-1-j} P_{2j-1}(u,v),\quad
\tg_{2k-3}(u,v,b)=\sum_{j=1}^{k-1} b^{k-1-j} \tP_{2j-1}(u,v)$$ So
the total number of mixed-type sections is $4\sum_{j=1}^{k-1}j.$
To see that these sections decompose as
$\sum_{j=1}^{k-1}\Bigl( V_{2j+1}+V_{2j-1}\Bigr)$ we write them  in a
unitary trivialization (in the chart $u\neq 0$)
$$\rho_{unit}=w^2f_{(2k)}\bigl(1+\vert z \vert^2\bigr)^{-k}
\Bigl(\bigl(z \tg_{2k-3}-g_{2k-3}\bigr)e_1-\bigl(\barz g_{2k-3}+\tg_{2k-3}\bigr)e_2
\Bigr).$$
where
$$\tg_{2k-3}=\sum_{j=1}^{k-1}w^{2k-2-2j}\sum_{n=0}^{2j-1}a^{(j)}_n z^{2j-1-n},\quad
g_{2k-3}=\sum_{j=1}^{k-1}w^{2k-2-2j}\sum_{n=0}^{2j-1}c^{(j)}_n z^{2j-1-n}$$
Then, $\rho_{unit}$ is brought to the form
$$\rho_{unit}=w^{2k}f_{(2k)}(\lambda)\bigl(1+\vert z\vert^2\bigr)^{-k}\Biggl(
e_1\sum_{j=1}^{k-1}w^{-2j}\Bigl(a^{(j)}_0z^{2j}-
c^{(j)}_{2j-1}+\sum_{n=0}^{2j}\beta^{(j)}_n z^{2j-n}\Bigr)+$$
$$(-){\barw\over w}e_2\sum_{j=1}^{k-1}{1\over w^{2j-2}\vert w\vert^{2}}
\Bigl(a^{(j)}_0z^{2j-1}+c^{(j)}_{2j-1}\barz+\sum_{n=1}^{2j-1}(a^{(j)}_n+
c^{(j)}_{n-1}\vert z \vert^2) z^{2j-1-n}\Bigr)\Biggr)$$
where
$$\beta^{(j)}_0=a^{(j)}_0,\quad \beta^{(j)}_{2j}=-c^{(j)}_{2j-1},\quad \beta^{(j)}_n=a^{(j)}_n-c^{(j)}_{n-1},\quad n=1,\ldots,2j-1
$$
Now recall that $e_1$ and ${\barw\over w}e_2$ are $SL(2,\CC)$ invariant (1,0) forms,
and $\lambda={\vert w \vert^2 \over 1+\vert z\vert^2}$ is $SL(2,\CC)$  invariant.
We see that the $e_1$ piece in $\rho_{unit}$ transforms as $\sum_{j=1}^{k-1}V_{2j+1},$ i.e.
for each $j=0,\ldots, k-1$
$$w^{-2j}\sum_{n=0}^{2j}\beta^{(j)}_n z^{2j-n}$$
transforms as $V_{2j+1}.$

The $e_2$ piece in $\rho_{unit}$  transforms as $\sum_{j=1}^{k-1}V_{2j-1}$ if
we impose $2j+1$ constraints  for each $j=0,\ldots,k-1$
$$a^{(j)}_0=0,\quad c^{(j)}_{2j-1}=0,\quad a^{(j)}_n=c^{(j)}_{n-1} \quad n=1,\ldots 2j-1.$$

All these sections are in $L^2.$ Indeed, the norm of each section
in $V_{2j+1}$ is not greater than
$${\pi \over 2}\int {d\lam \over \lam} f_2 \, f_{(2k)}^2
 \, \int {\vert z^{2j} w^{2(k-j)} \vert^{2} dz d\barz \over \bigl(1+\vert z\vert^2\bigr)^{2k+2}},$$
where $j=1,\ldots, k-1.$

Using (\ref{defii}) the integral is
brought to the form
$${\pi \over 2}\int {d\lam \over \lam} \lam^{2(k-j)}\, f_2\, f_{(2k)}^2
 \, \int {\vert z \vert^{4j} dz d\barz \over \bigl(1+\vert z\vert^2\bigr)^{2+2j}}, $$
which is finite in the relevant range, i.e. for $j=1,\ldots,k-1.$

Analogously, the norm of each section
in $V_{2j-1}$ is not greater than
$${\pi \over 2}\int {d\lam \over \lam} f_1 \, f_{(2k)}^2 \lam^{2(k-j)}
 \, \int {\vert z\vert^{2(2j-2)} dz d\barz \over \bigl(1+\vert z\vert^2\bigr)^{2j}},$$
which is finite in the relevant range, i.e. for $j=1,\ldots,k-1.$

\subsubsection{$H^0\bigl(T\PP^1,\Omega^2(2k)\bigr),\quad k>0.$} Finally we find
holomorphic sections of the line bundle $\Omega^2(2k)$ on $T\PP^1$.
In a  holomorphic trivialization they are
$$ \rho_{hol}=F_{2k-4}(u,v,b)\bigl(vdu-udv\bigr)\wdg \bigl(db-b{du\over u}-b{dv \over v}\bigr),$$
where
$$F_{2k-4}=\sum_{j=0}^{k-2}b^{k-2-j}\, P_{2j}(u,v).$$
In a unitary trivialization they have the form
$$\rho_{unit}=w^{2k}{f_{(2k)}(\lambda)\over \lambda}\bigl(1+\vert z \vert^2\bigr)^{-k}\,\sum_{j=0}^{k-2}
w^{-2j}\, \sum_{p=0}^{2j} a^{(j)}_p z^{2j-p}\,  e_1 \wdg \left(e_2
{\barw \over w}\right),$$ so we conclude that they transform in a
representation $\sum_{j=0}^{k-2}V_{2j+1}.$

All these sections have
finite $L^2$ norm, since for $j=0,\ldots,k-2$ and $p=0,\ldots, 2j$ we find
$${\pi \over 2}\int {d\lam \over \lam} \lam^{2(k-j-1)}\, f_{(2k)}^2
 \, \int {\vert z \vert^{2p} dz d\barz \over \bigl(1+\vert z\vert^2\bigr)^{2+2j}} < \infty.$$

\subsection{Wavefunctions on $T^*\PP^1.$}
We regard $T^*\PP^1$ as the total space of the line bundle $\cO(-2)$
over $\PP^1$ and use homogeneous coordinates $u,v,b'$ with $\CC^*$
weights $1,1,-2$. In the patch $u\neq 0$ we define inhomogeneous
coordinates \be z={v\over u},\quad w'=\sqrt{b'}u \label{newdefi} \ee
Our goal is to compute the $L^2$ Dolbeault cohomology of the bundles
$\Omega^i(-2m)$, $i=0,1,2$.

\if
To confirm the "bubbled"(i.e. with zero magnetic charge) part of
the basic OPE (\ref{basic}) we need to compute image of the Chern
map of the K-theory class
$$K\bigl(T^*P^1,O(2m)\bigr)=\sum_{p=0}^2\sum_{q=0}^2 (-)^{p+q}
H^p\bigl(\Omega^q(2m),T^*P^1\bigr)$$ where $\Omega^q(k)=\Omega^q
\otimes O(k).$
\fi

\subsubsection{Metrics}

The most general $SU(2)$-invariant K\"ahler  form on $T^*\PP^1$ is
$$J=f_1(x) e_1' \wdg \bare'_1+ f_2(x) e_2 \wdg \bare_2,$$
where \be e_1'={dw'\over w'}+\barz e_2, \quad e_2={dz\over 1+\vert z
\vert^2} \label{newdefii} \ee and $f_1,f_2$ are functions of THE
$SU(2)$ invariant \be x:=\vert w' \vert^2 (1+\vert z\vert^2).
\label{newdefiii} \ee From $dJ=0$ we find $f_1=x f_2'$ so that
geometry is specified in terms of a single function $f_2(x)$, which
we take to be a positive function with the following asymptotics:
\be f_2\rightarrow x \quad for \quad  x \rightarrow \infty,\quad
f_2\rightarrow const \quad for \quad x \rightarrow 0.
\label{newasymp} \ee The first condition ensures that at $x
\rightarrow \infty$ the metric becomes flat. The second condition is
required so that for $x=0$ the metric is nonsingular.

Next consider the line bundle $\cO(2k)$ over $T^*\PP^1$. In a
unitary trivialization the connection on this bundle is \be
 A^{(2k)}={x f'_{(2k)}\over f_{(2k)}}e_1'-k\barz e_2,\quad {\overline A}^{(2k)}=-{x f'_{(2k)}\over f_{(2k)}}\bare_1'+k z \bare_2
\label{newconn} \ee
and covariant differentials are defined as
$$ D=\p+A^{(2k)},\quad \Delbar=\delbar +{\overline A}^{(2k)}$$
For $k=-m,\, m\in \ZZ,$ the function  $f_{(-2m)}$ has the
asymptotics \be f_{(-2m)} \rightarrow x^{-m} \quad for \quad x
\rightarrow \infty,\quad f_{(-2m)}\rightarrow 1 \quad for \quad x
\rightarrow 0. \label{newasympii} \ee The behavior for $x
\rightarrow \infty$ is chosen in such a way that asymptotically the
holomorphic section $w'^{2m}$ of $\cO(-2m)$ has constant pointwise
norm. The reason for this choice is that $w'^{2m}$ continues in the
limit $x \rightarrow \infty$  to a section which transforms
trivially between the two charts $u\ne 0$ and $v\ne 0.$ The behavior
for $x \rightarrow 0$ ensures that we have a nonsingular metric when
restricting to the zero section of $T^*\PP^1$.

\subsubsection{A heuristic argument}

Again we begin with a heuristic argument. The sheaf $\cO(2)$ on
$T^*\PP^1$ can be identified with $\cO(-D)$, where $D$ is the zero
section $b'=0$. A short exact sequence of sheaves
\begin{equation}\label{shortEH}
0\ra \cO(-D)\ra \cO\ra \cO_D\ra 0
\end{equation}
implies a long exact sequence for sheaf cohomology. Let us assume
that it holds also for $L^2$ Dolbeault cohomology. We also recall
\cite{KW} that for $m=0$ the only square-integrable solution of
equations (\ref{eqsdolbope}) on $T^*\PP^1$ is of type $(1,1)$, so
$$
H^1(T^*\PP^1,\Omega^1)=V_1,
$$
and all other $L^2$ Hodge numbers on $T^*\PP^1$ vanish. Then the
long exact sequence coming from (\ref{shortEH}) and its relatives
obtained by tensoring (\ref{shortEH}) with $\Omega^i$ imply
$$
H^1(T^*\PP^1,\cO(2))=V_1,\quad H^1(T^*\PP^1,\Omega^2(2))=V_1,\quad
H^1(T^*\PP^1,\Omega^1(2))=V_3,
$$
and all other cohomologies for $m=-1$ vanish. Now that we know
cohomology for $m=-1$, we can tensor (\ref{shortEH}) with
$\Omega^i(2)$ and determine cohomology for $m=-2$, etc. In this way
we obtain the following predictions for dimensions of  $L^2$
cohomology groups for $k=-m>0$:
$$H^1\bigl(T^*\PP^1,\Omega^1(2k)\bigr)=V_1+V_{2k-1}+V_{2k+1}+
2\sum_{j=1}^{k-2}V_{2j+1},\quad k\ge 3,$$
$$ H^1\bigl(T^*\PP^1,\Omega^1(2)\bigr)=V_3,\quad H^1\bigl(T^*\PP^1,\Omega^1(4)\bigr)=V_1+V_3+V_5,$$
$$H^1\bigl(T^*\PP^1,\Omega^2(2k)\bigr)=H^1\bigl(T^*\PP^1,\cO(2k)\bigr)=
\sum_{j=0}^{k-1}V_{2j+1}.$$ The results for $k<0$ are obtained by
Kodaira-Serre duality; in fact, from the above formulas it is easy
to see that cohomology groups depend only on $|k|$.

Below we will find exactly the right number of square-integrable
solutions of (\ref{eqsdolbope}) in cohomological degree $1$, with
the correct transformation properties under $PSU(2)$. We also
checked that in degree zero (and by Kodaira-Serre duality, in degree
$2$) all $L^2$ cohomology vanishes, just as the long exact sequence
predicts. We have not been able to verify that we have found all
square-integrable solutions of (\ref{eqsdolbope}) in degree $1$. We
only checked that if other solutions in degree $1$ exist, they
cannot transform in $PSU(2)$ representations of dimensions $1$ and
$3$.

\subsubsection{$H^1\bigl(\Omega^1(2k)\bigr),\quad k>0.$} The most general ansatz
(in the unitary trivialization and in the chart $u\neq 0$) for the
component of the $(n+1)$-plet in $H^1\bigl(\Omega^1(2k)\bigr)$ with
the $PSU(2)$ isospin projection $J_3=-(n/2)$ is: \be
\omega={f_{(2k)}\over (1+ \vert z \vert^2)^{k}} w'^{-2k} \, \Biggl(a
e_1' \wdg \bare'_1 +be_2\wdg \bare_2+c{\barw'\over w'} e_1'\wdg
\bare_2+ d{w'\over \barw'} e_2\wdg \bare'_1\Biggr), \label{start}
\ee where
$$a=\sum_{p=0}^n \, a_n(x) w'^{n-p}\,(\barw' \barz)^p$$
and the functions $b,c,d$ have a similar form. We have used that
$e'_1$ and ${w'\over \barw'}e_2$ are $SU(2)$-invariant $(1,0)$
forms. Various terms in $a$ correspond to different ways of building
up the component of a $(n+1)$-plet with $J_3=-(n/2),$ i.e.
$u^n,u^{n-1}\barv,\ldots ,\barv^n$.

Imposing $D(*\omega)=0$ and $\Delbar(\omega)=0$ we found that
non-trivial cohomology groups come from using two simple special
cases of the general ansatz (\ref{start}).

${\bf I.}~~~$The first simplified ansatz has the form: \be
\omega={w'^{-2k} f_{(2k)}\over \bigl( 1 +\vert
z\vert^2\bigr)^{k}}\Omega_n, \label{ans} \ee where
$$\Omega_n=w'^n\Biggl(a(x) e_1' \wdg \bare'_1
+b(x) e_2\wdg \bare_2+d(x)\barz e_2\wdg \bare'_1\Biggr).$$ From
$\Delbar \omega=0$ we find \be a-xb'+d=0.\label{haha} \ee Meanwhile,
$D(*\omega)=0$ gives \be 2k{f_1\over f_2}b-xd'-2xd{f'_{(2k)}\over
f_{(2k)}} +(n-2k)\Bigl({f_1b\over f_2}-d\Bigr)=0, \label{hahaii} \ee
\be x\left({f_2 a\over f_1}\right)'-{f_1\over f_2}b+
\Bigl(n-2k+2x{f'_{(2k)}\over f_{(2k)}}\Bigr){f_2\over f_1}a=0.
\label{hahaiii} \ee

${\bullet ~~~}$Let us first assume $n=2k$, then a linear combination
of (\ref{hahaii}) and (\ref{hahaiii}) gives
$$\Bigl(2k{f_2\over f_1}a-d\Bigr)'+
2{f'_{(2k)}\over f_{(2k)}}\Bigl(2k{f_2\over f_1}a-d\Bigr)=0,$$ which
can be integrated to express $d$ in terms of $a$ as \be
d=2k{f_2\over f_1}a-{C_0\over f_{(2k)}^2} \label{uraii} \ee where
$C_0$ is an integration constant. {}From (\ref{hahaiii}) $b$ can
also be expressed in terms of $a$ and its derivative, so that the
system (\ref{haha}-\ref{hahaiii}) reduces to a second-order
inhomogeneous differential equation: \be -x^2 {f_2\over
f_1}\phi''-\left(x\left({xf_2\over f_1}\right)'+ 2x^2{f_2\over
f_1}{f'_{(2k)}\over f_{(2k)}}\right)\phi'+ \left(2k+{f_1\over f_2}-
2x\left({xf_2\over f_1}{f'_{(2k)}\over f_{(2k)}}\right)'\right)\phi=
{C_0\over f_{(2k)}^2}, \label{finii} \ee where $\phi={f_2 a \over
f_1}.$ Near $x \rightarrow \infty$ (\ref{finii}) becomes
$$-x^2\phi''-(1+2k)x\phi'+(1+2k)\phi={C_0\over x^{2k}},$$
and its general solution behaves at infinity as \be \phi={C_0\over
(1+2k)x^{2k}}+C_1x+{C_2\over x^{1+2k}} \label{infsolii} \ee where
$C_1$ and $C_2$ parameterize the general solution of the homogeneous
equation.

Near $x\rightarrow 0$ (\ref{finii}) becomes
$$-x\phi''+\phi'+2kx\phi=C_0 x,$$
and its general solution behaves at zero as \be \phi={C_0\over
2k}+C_3x^2+C_4x^2Logx, \label{zerosolii} \ee where $C_3$ and $C_4$
parameterize the general solution of the homogeneous equation.

To ensure that $\omega$ is well-behaved  near the origin we must
choose $C_4=0.$  This is always possible. Starting from any  decaying solution
of the inhomogeneous equation at infinity
$$\phi_{inhom}={C_0\over (1+2k)x^{2k}}+{\tC_2\over x^{2k+1}}$$
we may always add a decaying solution of the homogeneous equation so
that
$$\phi=\phi_{inhom}+{C_2\over x^{2k+1}}$$
continues to small $x$ in a desired way, i.e. $C_4=0.$

Finally we note that $\omega$  has a finite $L^2$ norm:
$$
 {\pi \over 2}\int {dx \over x} f_{(2k)}^2
 \, \int {dz d\barz \over
\bigl(1+\vert z\vert^2\bigr)^{2+2k}}\left(a^2 {f_2\over f_1}+
 b^2 {f_1\over f_2}+d^2\vert z \vert^2\right)$$
Indeed, using the asymptotics at $x\rightarrow \infty$
$$a \sim {1\over x^{2k}},\quad b \sim {1\over x^{2k}}, \quad
d \sim {1\over x^{2k}}$$ we find that integral converges for $2k \ge
2.$ We conclude that using ansatz (\ref{ans}) for $n=2k$ we found a
well-behaved $(2k+1)$-plet with finite $L^2$ norm.

$\bullet~~~~ n\le 2k-2,\, n>0$

Let us consider (\ref{haha}-\ref{hahaiii}) with $n\ne 2k.$ Then a
linear combination of (\ref{hahaii}) and (\ref{hahaiii}) can be
integrated to express $d$ in terms of $a$ as \be d=n{f_2\over
f_1}a-{C_0 x^{2k-n} \over f_{(2k)}^2}, \label{uraiii} \ee where
$C_0$ is an integration constant. From equation (\ref{hahaiii}) $b$
can also be expressed in terms of $a$ and its derivative, so that
the system (\ref{haha}-\ref{hahaiii}) reduces to a second-order
inhomogeneous differential equation: \be -x^2 {f_2\over f_1}\phi''-
\left(x\left({xf_2\over f_1}\right)'+ 2x^2{f_2\over
f_1}{f'_{(2k)}\over f_{(2k)}}+{xf_2(n-2k)\over f_1} \right)\phi'+
\label{finiii} \ee
$$\left(n+{f_1\over f_2}-
2x\left({xf_2\over f_1}{f'_{(2k)}\over f_{(2k)}}\right)'+
(2k-n)x\left({f_2\over f_1}\right)'\right)\phi= {C_0 x^{2k-n}\over
f_{(2k)}^2},$$ where $\phi={f_2 a \over f_1}.$ Near $x\rightarrow
\infty$ (\ref{finiii}) becomes
$$-x^2\phi''-(1+n)x\phi'+(1+n)\phi={C_0\over x^n},$$
and its general solution at infinity is \be \phi={C_0\over
(1+n)x^n}+C_1x+{C_2\over x^{1+n}} \label{infsoliii} \ee where $C_1$
and $C_2$ parameterize the general solution of the homogeneous
equation.

We must set $C_1=0$ to obtain $\omega$ with a finite $L^2$ norm for
$n>0$:
$$\int_{T^*\PP^1} \omega \wdg *{\overline  \omega}=
\int {dx \over x^{1+n}} \, \int {dz d\barz \over \left(1+\vert
z\vert^2\right)^{2+n}}.$$ Near $x\rightarrow 0$ (\ref{finiii})
becomes
$$-x^2\phi''+(2k-n+1)x\phi'+2(n-2k)\phi=C_0 x^{2k-n+2}$$
For $n < 2k-2$ general solution near zero is \be \phi={C_0
x^{2k-n+2}\over 2(n-2k)}+C_3x^2+C_4x^{2k-n}, \label{zerosoliii} \ee
and for $n=2k-2$ \be \phi=-{C_0 x^{4}\over 4}+C_3x^2+C_4 x^2 Log(x),
\label{zerosoliv} \ee where $C_3$ and $C_4$ parameterize the general
solution of the homogeneous equation.

For $n=2k-2$ there is a good solution if $C_4=0.$ For
even\footnote{Recall that $x^2$ is a good coordinate, but $x$ is
not, so odd powers of $x$ are ill-behaved.} $n$ such that $n<2k-2$
there is a good solution if $C_3=0$. Such solutions always exist.
Starting from any decaying solution of the inhomogeneous equation at infinity
$$\phi_{inhom}={C_0\over (1+n)x^n}+{\tC_2\over x^{n+1}}$$
we may always add a decaying solution of the homogeneous equation so
that
$$\phi=\phi_{inhom}+{C_2\over x^{n+1}}$$
continues to small $x$ in the desired way, i.e. $C_4=0$ or $C_3=0.$
We conclude that using the ansatz (\ref{ans}) we found a
well-behaved $(n+1)$-plet with a finite norm for even $n$ such that
$n>0$ and $n\le 2k-2.$

${\bf II.}~~~$The second  simplified ansatz has the form: \be
\omega={w'^{n-2k} f_{(2k)}\over \bigl( 1 +\vert z\vert^2\bigr)^{k}}
d(x){w'\over \barw'} e_2 \wdg \bare'_1. \label{ansnew} \ee Imposing
$D(*\omega)=0$ and $\Delbar(\omega)=0$ gives \be
d(x)={x^{2k-n-1}\over f_{(2k)}^2}. \label{solnew} \ee For $x
\rightarrow 0$  $\omega$ is well-behaved if $n$ is even and
satisfies the inequality $n\le 2k-4.$ Also, this solution has finite
$L^2$ norm for $n>0$:
$$\int \omega \wdg *\overline{\omega} \sim
 \int {dx \over x^{n+3}}
 \, \int {dz d\barz \over
\bigl(1+\vert z\vert^2\bigr)^{2+n}}.$$

\subsubsection{$H^1\bigl(\cO(2k)\bigr)$ and $H^1\bigl(\Omega^2(2k)),\quad k>0$}
For $k>0$ we start from an ansatz (in the unitary trivialization and
in the chart $u\neq 0$) for the component of the $(n+1)$-plet in
$H^1\bigl(O(2k)\bigr)$ with $J_3=-(n/2)$: \be \omega={f_{(2k)}\over
(1+ \vert z \vert^2)^{k}} w'^{n-2k} \, \Biggl(\beta(x) \bare'_1
+\alpha(x) {\barw'\over w'} \bare_2 \Biggr). \label{startferm} \ee

Imposing $\Delbar(\omega)=0$ and $D(*\omega)=0$ gives the following
result. For even $n$ such that $0\le n\le 2k-2$
$$\omega=w'^n \left({\barw'\over w'}\right)^{k}
{x^{k-n}\over f_{(2k)}f_2}\bare'_1$$ belongs to
$H^1\bigl(\cO(2k)\bigr),$ has finite $L^2$ norm and is well-behaved
for $x\rightarrow 0.$

The component of the $(n+1)$-plet in $H^1\bigl(\Omega^2(2k)\bigr)$
with $J_3=-(n/2)$ can be found analogously. For even $n$ such that
$0 \le n\le 2k-2$
$$\omega=w'^n \left({\barw'\over w'}\right)^{k-1}
{x^{k-n-1}f_1 \over f_{(2k)}}e_1'\wdg e_2\wdg  \bare'_1$$ belongs to
$H^1\bigl(\Omega^2(2k)\bigr),$ has finite $L^2$ norm and is
well-behaved for $x\rightarrow 0.$

\subsection{Testing S-duality}

We are now ready to perform a test of the S-duality prediction
(\ref{maineqtotest}). Summing up all cohomology groups
$H^p(T^*\PP^1,\Omega^q(-2m))$ with the sign $(-1)^{p+q}$, we find
the ``bubbled'' contribution to the zeromode Hilbert space:
$$
V_{2m+1}-V_1-V_{2m-1},
$$
where $V_{2j+1}$ is the $2j+1$-dimensional representation of
$PSL(2,\CC)$. This corresponds to the sum of Wilson loops
$$
W_{2m}-W_0-W_{2m-2},
$$
in precise agreement with the S-duality prediction
(\ref{maineqtotest}).

As a consistency check on our computation, let us consider the Euler
characteristics of the graded vector spaces $\frV(\sA,\sB,\ldots)$
associated to the the left-hand side and right-hand side of eq.
(\ref{eqsdolbope}). According to our computations, the ``bulk''
contribution to the Euler characteristic of the right-hand side is
$$
1+(2m+1)-(2m-1)=3.
$$
The ``bubbled'' contribution is
$$
(2m+1)-1-(2m-1)=1.
$$
Therefore the Euler characteristic of the right-hand side is $4$. We
can compute the Euler characteristic of the left-hand side by moving
the WH operators so that they are inserted at the same point on the
interval $I$ but at different points on $C$.\footnote{Unlike in
\cite{KW}, there is no natural flat connection on the sheaf of the
zeromode Hilbert spaces $\frV(\sA,\sB,\ldots)$, and in principle the
stalk of this sheaf might depend on the locations of the insertion
points. Nevertheless, while the dimensions of the individual graded
components might jump, the Euler characteristic must be constant.}
If the WH operators are inserted at different points on $C$, the
space of zero modes factorizes, and so does the Euler
characteristic. The Hilbert space $\frV(WT_{1,0})$ is purely even
and two-dimensional. The Hilbert space $\frV(WT_{1,2m})$ is
$$
\oplus_{p,q=0}^1 H^p(\PP^1,\Omega^q(-2m)),
$$
and its Euler characteristic is $2$ for any $m$. Therefore the Euler
characteristic of the left-hand side is also $4$.

\section{Concluding remarks}

As mentioned in the introduction, 't Hooft line operators can be
interpreted mathematically as objects of the category of equivariant
perverse sheaves on the affine Grassmannian $Gr_G$. Then the algebra
of loop operators can be identified with the K-theory of this
category, and the S-duality prediction is equivalent to the
geometric Satake correspondence. 't Hooft loop operators labeled by
coweights of $G$ define a distinguished basis in the $K^0$-group.

It was suggested by R.~Bezrukavnikov that the algebra of Wilson-'t
Hooft loop operators can be similarly interpreted as the $K^0$-group
of the equivariant derived category of coherent sheaves on a certain
subset $\Lambda_G$ of the cotangent bundle of $Gr_G$. $\Lambda_G$ is
defined as the union of the conormal bundles to the Schubert cells
in $Gr_G$ and is invariant under the left $G[[z]]$ action on $Gr_G$.
Just like $Gr_G$ parameterizes Hecke transformations of holomorphic
$G$-bundles, $\Lambda_G$ parameterizes Hecke transformations of
Higgs bundles. Thus any object of the $G[[z]]$-equivariant derived
category of $\Lambda_G$ can be used to define a functor from the
derived category of $\cM_{Higgs}(G,C)$ to itself and can be thought
of as a line operator. It was proved in \cite{BFM} that the
$K^0$-group of $D^b_{eq}(\Lambda_G)$ is the Weyl-invariant part of
the group algebra of $\dL(G)$, in agreement with the physical
arguments. Further, it was conjectured in \cite{BFM} that the
obvious invariance of $\dL(G)$ under the exchange of $G$ and $\LG$
comes from an equivalence between categories $D^b_{eq}(\Lambda_G)$
and $D^b_{eq}(\Lambda_\LG)$. From the physical viewpoint, this
conjecture means that the categories of line operators for $G$ and
$\LG$ are equivalent and thus follows from the S-duality conjecture.

Note also that the physical definition of the Wilson-'t Hooft loop
operator suggests that there is a distinguished basis in the
K-theory of $D^b_{eq}(\Lambda_G)$ labeled by elements of
$\dL(G)/\cW$, and that the S-duality group acts on this basis in a
natural way. The mathematical significance of this basis remains
unclear. Moreover, Wilson-'t Hooft {\it line} operators should
correspond to some distinguished objects in $D^b_{eq}(\Lambda_G).$
It was conjectured by R. Bezrukavnikov that these distinguished
objects are certain perverse coherent sheaves on $\Lambda_G$.

\section*{Acknowledgments}

We would like to thank R.~Bezrukavnikov, A.~Braverman, S.~Gukov,
M.~Finkelberg, I.~Mirkovic, L.~Positselski and E.~Witten for
discussions. We are especially grateful to R.~Bezrukavnikov for
valuable advice without which this work would not be possible. We
would like to express our thanks to the Aspen Center for Physics for
hospitality. A.K. is also grateful to the Independent University of
Moscow for staying open during the winter holidays of 2006-2007 and
thereby providing an opportunity to share some preliminary results
with interested mathematicians and to receive their feedback. This
work was supported in part by the DOE grant DE-FG03-92-ER40701.

\end{document}